\documentclass[acmsmall]{acmart}
\usepackage[inline]{enumitem}

\usepackage{booktabs} 
\usepackage{pgf}
\usepackage{colortbl}
\usepackage{makecell}

\usepackage{array}
\usepackage{wasysym}
\newcommand*\rot{\rotatebox{90}}
\usepackage{graphicx}
\usepackage{subfig}

\usepackage{varwidth}
\usepackage{paralist}
\usepackage{inputenc}
\usepackage{multirow}
\usepackage{xcolor}
\usepackage{geometry}
\usepackage{soul}
\usepackage{pdflscape}
\usepackage{algorithm, algpseudocode}
\def\BibTeX{{\rm B\kern-.05em{\sc i\kern-.025em b}\kern-.08emT\kern-.1667em\lower.7ex\hbox{E}\kern-.125emX}}

\newcommand*{\totaldpinst}{1,818}
\newcommand*{\totaldpweb}{1,254}
\newcommand*{\totaldpperc}{11.1\%}
\newcommand*{\totaldeci}{234}
\newcommand*{\totaldecw}{183}

\newcommand*{\fai}{6}
\newcommand*{\faw}{6}

\newcommand*{\mdw}{244}

\newcommand*{\scarw}{609}

\newcommand*{\sneakw}{23}

\newcommand*{\spw}{275}

\newcommand*{\urgw}{437}

\newcommand*{\ani}{313}
\newcommand*{\anw}{264}

\newcommand*{\anid}{29}
\newcommand*{\anwd}{20}

\newcommand*{\csi}{169}
\newcommand*{\csw}{164}

\newcommand*{\cti}{393}
\newcommand*{\ctid}{157}
\newcommand*{\ctw}{361}
\newcommand*{\ctwd}{140}

\newcommand*{\few}{6}

\newcommand*{\htci}{31}
\newcommand*{\htcw}{31}

\newcommand*{\hci}{5}
\newcommand*{\hcw}{5}

\newcommand*{\hsi}{14}
\newcommand*{\hsw}{13}

\newcommand*{\hdi}{47}
\newcommand*{\hdw}{43}

\newcommand*{\lti}{88}
\newcommand*{\ltw}{84}

\newcommand*{\lsi}{632}
\newcommand*{\lsw}{581}
\newcommand*{\lsdi}{17}
\newcommand*{\lsdw}{17}

\newcommand*{\psei}{67}
\newcommand*{\psew}{62}

\newcommand*{\sibi}{7}
\newcommand*{\sibw}{7}

\newcommand*{\testi}{12}
\newcommand*{\testw}{12}

\newcommand*{\tqi}{9}
\newcommand*{\tqw}{9}

\newcommand*{\vii}{25}
\newcommand*{\viw}{24}

\newcommand{\ReturnNewLine}{\State\Return}
\begin{document}
\title[Dark Patterns at Scale]{Dark Patterns at Scale: Findings from a Crawl of 11K Shopping Websites}

\setcopyright{acmlicensed}
\acmJournal{PACMHCI}
\acmYear{2019} \acmVolume{3} \acmNumber{CSCW} \acmArticle{81} \acmMonth{11} \acmPrice{15.00}\acmDOI{10.1145/3359183}



\author{Arunesh Mathur}
\affiliation{%
  \institution{Princeton University}
  \streetaddress{304 Sherrerd Hall}
  \city{Princeton}
  \state{NJ}
  \postcode{08544}
  \country{USA}}
\email{amathur@cs.princeton.edu}
\author{Gunes Acar}
\affiliation{%
  \institution{Princeton University}
  \streetaddress{320 Sherrerd Hall}
  \city{Princeton}
  \state{NJ}
  \postcode{08544}
  \country{USA}}
\email{gunes@princeton.edu}
\author{Michael J.\ Friedman}
\affiliation{%
  \institution{Princeton University}
  \streetaddress{35 Olden Street}
  \city{Princeton}
  \state{NJ}
  \postcode{08544}
  \country{USA}}
\email{mjf4@princeton.edu}
\author{Elena Lucherini}
\affiliation{%
  \institution{Princeton University}
  \streetaddress{312 Sherrerd Hall}
  \city{Princeton}
  \state{NJ}
  \postcode{08544}
  \country{USA}}
\email{elucherini@cs.princeton.edu}
\author{Jonathan Mayer}
\affiliation{%
  \institution{Princeton University}
  \streetaddress{307 Sherrerd Hall}
  \city{Princeton}
  \state{NJ}
  \postcode{08544}
  \country{USA}}
\email{jonathan.mayer@princeton.edu}
\author{Marshini Chetty}
\affiliation{%
  \institution{University of Chicago}
  \streetaddress{355 John Crerar Library}
  \city{Chicago}
  \state{IL}
  \postcode{60637}
  \country{USA}}
\email{marshini@uchicago.edu}
\author{Arvind Narayanan}
\affiliation{%
  \institution{Princeton University}
  \streetaddress{308 Sherrerd Hall}
  \city{Princeton}
  \state{NJ}
  \postcode{08544}
  \country{USA}}
\email{arvindn@cs.princeton.edu}

\begin{abstract}
Dark patterns are user interface design choices that benefit an online service by coercing, steering, or deceiving users into making unintended and potentially harmful decisions. We present automated techniques that enable experts to identify dark patterns on a large set of websites. Using these techniques, we study shopping websites, which often use dark patterns to influence users into making more purchases or disclosing more information than they would otherwise. Analyzing $\sim$53K product pages from $\sim$11K shopping websites, we discover \totaldpinst{} dark pattern instances, together representing 15 types and 7 broader categories. We examine these dark patterns for deceptive practices, and find \totaldecw{} websites that engage in such practices. We also uncover 22 third-party entities that offer dark patterns as a turnkey solution. Finally, we develop a taxonomy of dark pattern characteristics that describes the underlying influence of the dark patterns and their potential harm on user decision-making. Based on our findings, we make recommendations for stakeholders including researchers and regulators to study, mitigate, and minimize the use of these patterns.

\end{abstract}

%
%
%
%



\maketitle

\renewcommand{\shortauthors}{Arunesh Mathur et al.}

\section{Introduction}
\label{section:intro}
Dark patterns~\cite{brignull-dark-patterns,gray-dark-patterns-2018} are user interface design choices that benefit an online service by coercing, steering, or deceiving users into making decisions that, if fully informed and capable of selecting alternatives, they might not make. Such interface design is an increasingly common occurrence on digital platforms including social media websites~\cite{frobrukerradet-deceived-2018}, shopping websites~\cite{brignull-dark-patterns}, mobile apps~\cite{bosch-tales-2016,app-dp}, and video games~\cite{zagal2013dark}. At best, dark patterns annoy and frustrate users. At worst, they can mislead and deceive users, e.g., by causing financial loss~\cite{affinion1,affinion2}, tricking users into giving up vast amounts of personal data~\cite{frobrukerradet-deceived-2018}, or inducing compulsive and addictive behavior in adults~\cite{schull2014addiction} and children~\cite{detour-act}. 

While prior work~\cite{conti2010malicious,gray-dark-patterns-2018,bosch-tales-2016,brignull-dark-patterns} has provided taxonomies to describe the existing types of dark patterns, there is no large-scale evidence documenting their prevalence, or a systematic and descriptive investigation of how the different types of dark patterns harm users. Collecting this information would allow us to first examine where, how often, and the technical means by which dark patterns appear; second, it would allow us to compare and contrast how various dark patterns influence users. In doing so, we can develop countermeasures against dark patterns to both inform users and protect them from such patterns. Further, given that many of these patterns are potentially unlawful, we can also aid regulatory agencies in addressing and mitigating their use.

In this paper, we present an automated approach that enables experts to identify dark patterns at scale on the web. Our approach relies on  \begin{enumerate*} \item a web crawler, built on top of OpenWPM~\cite{englehardt-tracking-2017,acar-tracking-2014}---a web privacy measurement platform---to simulate a user browsing experience and identify user interface elements; \item text clustering to extract all user interface designs from the resulting data; and \item inspecting the resulting clusters for instances of dark patterns. \end{enumerate*} We also develop a taxonomy so that researchers can share descriptive and comparative terminology to explain how dark patterns subvert user decision-making and lead to harm. We base this taxonomy on the characteristics of dark patterns as well as the cognitive biases they exploit in users.

While our automated approach generalizes, we focus this study on shopping websites, which are used by an overwhelming majority of people worldwide~\cite{pshopping}. Dark patterns found on these websites trick users into signing up for recurring subscriptions and making unwanted purchases, resulting in concrete financial loss. We use our web crawler to visit the $\sim$11K most popular shopping websites worldwide, create a large data set of dark patterns, and document their prevalence. Our data set contains several new instances and variations of previously documented dark patterns~\cite{brignull-dark-patterns,gray-dark-patterns-2018}. Finally, we use our taxonomy of dark pattern characteristics to classify and describe the patterns we discover. We have five main findings:

\begin{compactitem}
    \item We discovered \totaldpinst{} instances of dark patterns on shopping websites, which together represent 15 types of dark patterns and 7 broad categories.
    \item These \totaldpinst{} dark patterns were found on \totaldpweb{} of the $\sim$11K shopping websites ($\sim$\totaldpperc{}) in our data set. Shopping websites that were more popular, according to Alexa rankings \cite{alexa-top-sites-api}, were more likely to feature dark patterns. These numbers represent a lower bound on the total number of dark patterns on these websites, since our automated approach only examined text-based user interfaces on a sample of product pages per website.
    \item In using our taxonomy to classify the dark patterns in our data set, we discovered that the majority are \emph{covert}, \emph{deceptive}, and \emph{information hiding} in nature. Further, many patterns exploit cognitive biases, such as the default and framing effects. These characteristics and biases collectively describe the consumer psychology underpinnings of the dark patterns we identified.
    \item We uncovered \totaldeci{} instances of dark patterns---across \totaldecw{} websites---that exhibit deceptive behavior. We highlight the types of dark patterns we encountered that rely on deception. 
    \item We identified 22 third-party entities that provide shopping websites with the ability to create and implement dark patterns on their sites. Two of these entities openly advertised practices that enable deceptive messages.
\end{compactitem}

\vspace{1mm}
Through this study, we make the following contributions:
\begin{compactitem}
    \item We contribute automated measurement techniques that enable expert analysts to discover new or revisit existing instances of dark patterns on the web. As part of this contribution, we make our web crawler and associated technical artifacts available on GitHub\footnote{https://github.com/aruneshmathur/dark-patterns}. These can be used to conduct longitudinal measurements on shopping websites or be re-purposed for use on other types of websites (e.g., travel and ticket booking websites).
    \item We create a data set and measure the prevalence of dark patterns on 11K shopping websites. We make this data set of dark patterns and our automated techniques publicly available\footnote{https://webtransparency.cs.princeton.edu/dark-patterns} to help researchers, journalists, and regulators raise awareness of dark patterns~\cite{detour-act}, and to help develop user-facing tools to combat these patterns.
    \item We contribute a novel descriptive taxonomy that provides precise terminology to characterize how each dark pattern works. This taxonomy can aid researchers and regulators to better understand and compare the underlying influence and harmful effects of dark patterns. 
    \item We document the third-party entities that enable dark patterns on websites. This list of third parties can be used by existing tracker and ad-blocking extensions (e.g., Ghostery,\footnote{https://ghostery.com} Adblock Plus\footnote{https://adblockplus.com}) to limit their use on websites.
\end{compactitem}

\section{Related Work}
\label{section:background-and-related-work}

\subsection{Online Shopping and Influencing User Behavior}
Starting with Hanson and Kysar, numerous scholars have examined how companies abuse users' cognitive limitations and biases for profit, a practice they call market manipulation~\cite{hanson1999taking}. For instance, studies have shown that users make different decisions from the same information based on how it is framed~\cite{tversky-rational-choice-1989,tversky-framing-1981}, giving readily accessible information greater weight~\cite{tversky-judgment-1974}, and becoming susceptible to impulsively changing their decision the longer the reward from their decision is delayed~\cite{ainslie-specious-reward-1975}. Some argue that because users are not always capable of acting in their own best interests, some forms of `paternalism'---a term referring to the regulation or curation of the user's options---may be acceptable~\cite{thaler-libertarian-paternalism-2003}. However, determining the kinds of curation that are acceptable is less straightforward, particularly without documenting the practices that already exist.

More recently, Calo has argued that market manipulation is exacerbated by digital marketplaces since they posses capabilities that increase the chance of user harm culminating in financial loss, loss of privacy, and the ability to make independent decisions~\cite{calo2013digital}. For example, unlike brick-and-mortar stores, digital marketplaces can capture and retain user behavior information, design and mediate user interaction, and proactively reach out to users. Other studies have suggested that certain elements in shopping websites can influence impulse buying behavior~\cite{luo2018online, zhang2015mining}. For instance, perceived scarcity, social influence (e.g., `social proof'---informing users of others' behavior---and shopping with others~\cite{browne-ecommerce-2017, luo-shopping-2005}) can all lead to higher spending. More recently, Moser et al.\ conducted a study ~\cite{moser2019impulse} to measure the prevalence of elements that encourage impulse buying. They identified 64 such elements---e.g., product reviews/ratings, discounts, and quick add-to cart buttons---by manually scraping 200 shopping websites.

\subsection{Dark Patterns in User Interface Design}
\label{section:dark-patterns}

Coined by Brignull in 2010, dark patterns is a catch-all term for how user interface design can be used to adversely influence users and their decision-making abilities. Brignull described dark patterns as `tricks used in websites and apps that make you buy or sign up for things that you didn't mean to', and he created a taxonomy of dark patterns using examples from shopping and travel websites to help raise user awareness. The taxonomy documented patterns such as `Bait and Switch' (the user sets out to do one thing, but a different, undesirable thing happens instead), and `Confirmshaming' (using shame tactics to steer the user into making a choice).

\subsubsection{Dark Pattern Taxonomies} 
\label{sec:taxonomy}
A growing number of studies have expanded on Brignull's original taxonomy more systematically to advance our understanding of dark patterns. Conti and Sobiesk~\cite{conti2010malicious} were the first to create a taxonomy of malicious interface design techniques, which they defined as interfaces that manipulate, exploit, or attack users. While their taxonomy contains no examples and details on how the authors created the taxonomy are limited, it contains several categories that overlap with Brignull's dark patterns, including `Confusion' (asking the user questions or providing information that they do not understand) and `Obfuscation' (hiding desired information and interface elements). More recently, B{\"o}sch et al.\ ~\cite{bosch-tales-2016} presented a similar, alternative breakdown of privacy-specific dark patterns as `Dark Strategies', uncovering new patterns: `Forced Registration' (requiring account registration to access some functionality) and `Hidden Legalese Stipulations' (hiding malicious information in lengthy terms and conditions). Finally, Gray et al.\ ~\cite{gray-dark-patterns-2018} presented a broader categorization of Brignull's taxonomy and collapsed many patterns into categories such as `Nagging' (repeatedly making the same request to the user) and `Obstruction' (preventing the user from accessing functionality).

While these taxonomies have focused on the web, researchers have also begun to examine dark patterns in specific application domains. For instance, Lewis~\cite{lewis-irresistible-apps-2014} analyzed design patterns in the context of web and mobile applications and games, and codified those patterns that have been successful in making apps `irresistible', such as `Pay To Skip' (in-app purchases that skip levels of a game). In another instance, Greenberg et al.\ ~\cite{greenberg-proxemics-2014} analyzed dark patterns and `antipatterns'---interface designs with unintentional side-effects on user behavior---that leverage users' spatial relationship with digital devices. They introduced patterns such as `Captive Audience' (inserting unrelated activities such as an advertisement during users' daily activities) and `Attention Grabber' (visual effects that compete for users' attention). Finally, Mathur et al.\ ~\cite{Mathuram} discovered that most affiliate marketing on social media platforms such as YouTube and Pinterest is not disclosed to users (the `Disguised Ads' dark pattern).

\subsubsection{Dark Patterns and User Decision-making} A growing body of work has drawn connections between dark patterns and various theories of human decision-making in an attempt to explain how dark patterns work and cause harm to users. Xiao and Benbasat \cite{xiao-product-deception-2011} proposed a theoretical model for how users are affected by deceptive marketing practices in online shopping, including affective mechanisms (psychological or emotional motivations) and cognitive mechanisms (perceptions about a product). In another instance, B{\"o}sch et al.\ ~\cite{bosch-tales-2016} used Kahneman's Dual process theory~\cite{tversky-judgment-1974} which describes how humans have two modes of thinking---`System 1' (unconscious, automatic, possibly less rational) and `System 2' (conscious, rational)---and noted how `Dark Strategies' exploit users' System 1 thinking to get them to make a decision desired by the designer. Lastly, Lewis \cite{lewis-irresistible-apps-2014} linked each of the dark patterns described in his book to Reiss's Desires, a popular theory of psychological motivators~\cite{reiss-16-desires-2004}. Finally, a recent study by the Norwegian Consumer Council (Frobrukerr\r{a}det)~\cite{frobrukerradet-deceived-2018} examined how interface designs on Google, Facebook, and Windows 10 make it hard for users to exercise privacy-friendly options. The study highlighted the default options and framing statements that enable such dark patterns.

\subsection{Comparison to Prior Work}

Our study differs from prior work in two ways. First, while prior work has largely focused on creating taxonomies of the types of dark patterns either based on anecdotal data~\cite{brignull-dark-patterns,bosch-tales-2016} or data collected from users' submissions~\cite{gray-dark-patterns-2018,conti2010malicious}, we provide large-scale evidence documenting the presence and prevalence of dark patterns in the wild. Automated measurements of this kind have proven useful in discovering various privacy and security issues on the web---including third-party tracking \cite{englehardt-tracking-2017,acar-tracking-2014} and detecting vulnerabilities of remote third-party JavaScript libraries \cite{nikiforakis-third-party-js-2012}---by documenting how and on which websites these issues manifest, thus enabling practical solutions to counter them. Second, we expand on the insight offered by prior work about how dark patterns affect users. We develop a comprehensive taxonomy of dark pattern characteristics (Section~\ref{section:definitions}) that concretely explains the underlying influence and harmful effects of each dark pattern.

Finally, while prior work has shed light on impulse buying on shopping websites, the focus of our work is on dark patterns. While there is some overlap between certain types of dark patterns and impulse buying features of shopping websites~\cite{moser2019impulse}, the majority of impulse buying elements are not dark patterns. For instance, offering returns and exchanges for products, or showing multiple images of a product~\cite{moser2019impulse} do not constitute dark patterns: even though they play a role in persuading users into purchasing products, they do not fundamentally subvert user decision-making in a manner that benefits shopping websites and retailers.

\section{A Taxonomy of Dark Pattern Characteristics}
\label{section:definitions}


Our taxonomy explains how dark patterns affects user decision-making based on their characteristics as well as the cognitive biases in users---deviations from rational behavior justified by some `biased' line of reasoning~\cite{haselton2015evolution}---they exploit to their advantage. We ground this taxonomy in the literature on online manipulation~\cite{susser-online-manipulation, calo2013digital, wilkinson2013nudging} and by studying the types of dark patterns highlighted in previous work~\cite{brignull-dark-patterns,gray-dark-patterns-2018}. Our taxonomy consists of the following five dimensions:

\begin{compactitem}
  \item \textbf{Asymmetric}: Does the user interface design impose unequal weights or burdens on the available choices presented to the user in the interface?\footnote{We narrow the scope of asymmetry to only refer to explicit choices in the interface.} For instance, a website may present a prominent button to accept cookies on the web but make the opt-out button less visible, or even hide it in another page.
  
  \item \textbf{Covert}: Is the effect of the user interface design choice hidden from users? That is, does the interface design to steer users into making specific purchases without their knowledge? For instance, a website may leverage the decoy effect~\cite{decoy} cognitive bias, in which an additional choice---the decoy---is introduced to make certain other choices seem more appealing. Users may fail to recognize the decoy's presence is merely to influence their decision making, making its effect covert.
  
  \item \textbf{Deceptive}: Does the user interface design induce false beliefs either through affirmative misstatements, misleading statements, or omissions? For instance, a website may offer a discount to users that appears to be limited-time, but actually repeats when the user refreshes the website's page. Users may be aware that the website is trying to offer them a discount; however, they may not realize that they do not have a limited time to take advantage of the deal. This false belief affects users' decision-making i.e., they may act differently if they knew that the sale is recurring.
  
  \item \textbf{Hides Information}: Does the user interface obscure or delay the presentation of necessary information to the user? For instance, a website may not disclose additional charges for a product to the user until the very end of their checkout.
  
  \item \textbf{Restrictive}: Does the user interface restrict the set of choices available to users? For instance, a website may only allow users to sign up for an account with existing social media accounts so they can gather more information about them.
  
\end{compactitem}

Many types of dark patterns operate by exploiting cognitive biases in users. In Section \ref{section:findings}, we draw an explicit connection between each type of dark pattern we encounter and the cognitive biases it exploits. The biases we refer to in our findings are:
\begin{compactenum}
  \item Anchoring Effect~\cite{tversky-judgment-1974}: The tendency of individuals to overly rely on an initial piece of information---the `anchor'---in future decisions.
  \item Bandwagon Effect~\cite{sherif1936psychology}: The tendency of individuals to value something more because others seem to value it.
  \item Default Effect~\cite{johnson-defaults-2002}: The tendency of individuals to stick with options that are assigned to them by default due to inertia.
  \item Framing Effect~\cite{tversky-framing-1981}: The tendency of individuals to reach different decisions from the same information depending on how it is presented.
  \item Scarcity Bias~\cite{mittone2009scarcity}: The tendency of individuals to place a higher value on things that are scarce.
  \item Sunk Cost Fallacy~\cite{arkes1999sunk}: The tendency of individuals to continue an action if they have invested resources into it, even if that action might make them worse off.
 
\end{compactenum}

\section{Method}
\label{section:method}
Dark patterns may manifest in several different locations inside websites, and they can rely heavily upon interface manipulation, such as changing the hierarchy of interface elements or prioritizing certain options over others using different colors. However, many dark patterns are often present on users' \emph{primary interaction paths} in an online service or website (e.g., when purchasing a product on a shopping website, or when a game is paused after a level is completed). Further, multiple instances of a type of dark pattern share common traits such as the text they display (e.g., in the `Confirmshaming' dark pattern---which tries to shame the user into making a particular choice---many messages begin with \emph{No thanks}). Our technique relies on automating the primary interaction path of websites, extracting textual interface elements present in this path, and finally, grouping and organizing these---using clustering---for an expert analyst to sift through.

\begin{figure}[t]
\centering
\resizebox{0.9\textwidth}{!}{%
\frame{\includegraphics[width=\linewidth]{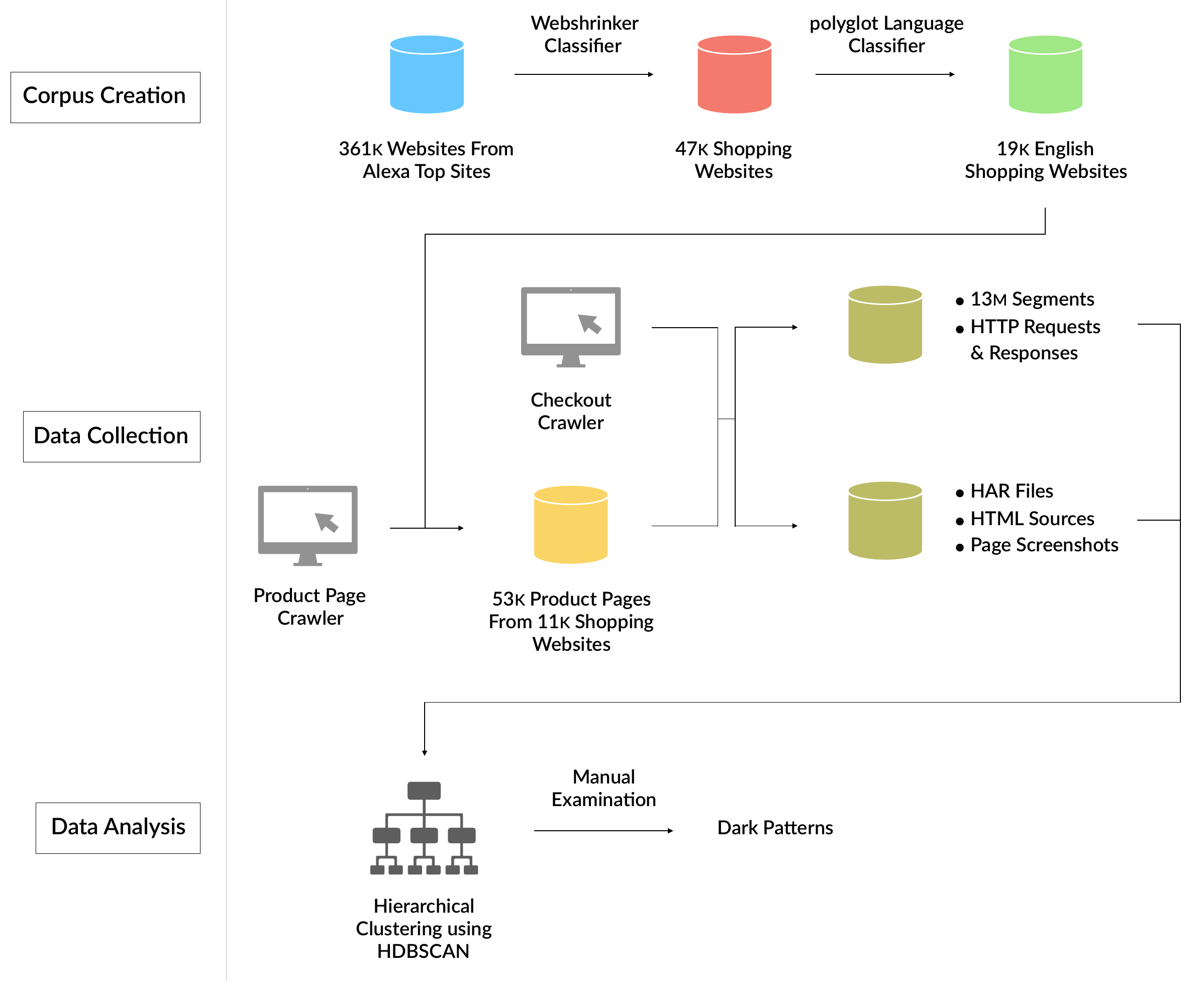}}}
\caption{Overview of the shopping website corpus creation, data collection using crawling, and data analysis using hierarchical clustering stages.}
\label{fig:method-diagram}
\end{figure}

While our method generalizes to different types of websites, we focus on shopping websites in this study. We designed a web crawler capable of navigating users' primary interaction path on shopping websites: making a product purchase. Our crawler aligned closely with how an ordinary user would browse and make purchases on shopping websites: discover pages containing products on a website, add these products to the cart, and check out. We describe these steps, and the data we collected during each visit to a website below. Figure \ref{fig:method-diagram} illustrates an overview of our method.

We note that only analyzing textual information in this manner restricts the set of dark patterns we can discover, making our findings a lower bound on the dark patterns employed by shopping websites. We leave detecting other kinds of dark patterns---those that are enabled using style, color, and other non-textual features---to future work, and we discuss possible approaches in Section~\ref{section:discussion}.

\subsection{Creating a Corpus of Shopping Websites}
\label{section:corpus}

We used the following criteria to evaluate existing lists of popular shopping websites, and, eventually, construct our own: \begin{enumerate*}
  \item the list must be representative of the most popular shopping websites globally, and
  \item the list must consist of shopping websites in English so that we would have the means to analyze the data collected from the websites.
\end{enumerate*}

We retrieved a list of popular websites worldwide from Alexa using the Top Sites API~\cite{alexa-top-sites-api}. Alexa is a web traffic analysis company that ranks and categorizes websites based on statistics it collects from users of its toolbar. We used the Top Sites list because it is more stable and is based on monthly traffic and not daily rank, which fluctuates often~\cite{lists-narseo} The list contained 361,102 websites in total, ordered by popularity rank.\footnote{We did not use Alexa's list of Top/Shopping websites~\cite{alexa-top-shopping} because of two issues. First, its criteria of categorization are not fully disclosed. Second, most of the websites in the list had an average monthly rank $>$ 500,000, which we did not consider to be representative of the most popular websites worldwide.}

We evaluated two website classification services to extract shopping websites from this list of the most popular websites: Alexa Web Information Service~\cite{awis} and WebShrinker~\cite{webshrinker}. We evaluated the classification accuracy of these services using a random sample of 500 websites from our list of 361K websites, which we manually labeled as `shopping' or `not shopping'. We considered a website to be a shopping website if it was offering a product for purchase. Of the 500 websites in our sample, we labeled 57 as `shopping' and 443 as `not shopping'. We then evaluated the performance of both classifiers against this ground truth.

Table~\ref{tab:confusion-matrices} in the Appendix summarizes the classifiers' results. Compared to Webshrinker, Alexa's classifications performed poorly on our sample of websites (classification accuracy: 89\% vs.\ 94\%), with a strikingly high false negative rate (93\% vs. 18\%). Although Webshrinker had a slightly higher false positive rate (0.2\% vs.\ 0.4\%), we used methods to determine and remove these false positives as we describe in Section~\ref{section:producturl}. 

We subsequently used Webshrinker to classify our list of 361K websites, obtaining a list of 46,569 shopping websites. To filter out non-English websites, we downloaded home pages of each site using Selenium~\cite{selenium} and ran language detection on texts extracted from the pages using the \texttt{polyglot} Python library \cite{py-polyglot}. Our final data set contained 19,455 English language shopping websites. We created this filtered list in August 2018.

\subsection{Data Collection with a Website Crawl}
\label{section:collection}

We conducted all our crawls from the Princeton University campus using two off-the-shelf computers, both equipped with 16G of memory and quad-core CPUs. Our crawler's exploration of each shopping website mimicked a typical user's primary interaction path on a shopping website---starting with one of its product pages. Therefore, the first step in our website crawl was to determine ways to automatically identify product URLs from shopping websites.

\subsubsection{Discovering Product URLs on Shopping Websites}
\label{section:producturl}

To effectively extract product URLs from shopping websites, we iteratively designed and built a Selenium-based web crawler that contained a classifier capable of distinguishing product URLs from non-product URLs. 

At first, we build a na\"ive depth-first crawler that, upon visiting a website's home page, determined the various URLs on the page, selected one URL at random, and then repeated this process from the selected URL. Using this crawler, we assembled a data set of several thousand URLs from visiting a random sample of 100 websites from our data set of 19K shopping websites. We manually labeled a sample of these URLs either as `product' or `non-product' URLs, and created a balanced data set containing 714 labeled URLs in total.

We trained a Logistic Regression classifier on this data set of labeled URLs using the \texttt{SGDClassifier} class from scikit-learn~\cite{scikit-learn}. We extracted several relevant features from this data set of URLs, including the length of a URL, the length of its path, the number of forward slashes and hyphens in its path, and whether its path contained the words `product' or `category'. We used 90\% of the URLs for training and obtained an 83\% average classification accuracy using five-fold cross validation.

We embedded this classifier into our original Selenium-based web crawler to help guide its crawl. As a result, rather than selecting and visiting URLs at random, the crawler first used the classifier to rank the URLs on a page by likelihood of being product URLs, and then visited the URL with the highest likelihood. The crawler declared a URL as product if its page contained an `Add to cart' or similar button. We detected this button by assigning a weighted score to visible HTML elements on a page based on their size, color, and whether they matched certain regular expressions (e.g., `Add to bag|cart|tote|\ldots'). This check also helped us weed out any false positives that may have resulted from the classification of shopping websites using Webshrinker (Section \ref{section:corpus}).

We tuned the crawler's search process to keep its crawl tractable. The crawler returned to the home page after flagging a product URL. It did not visit a given URL more than two times to avoid exploring the same URLs, and it stopped after visiting 100 URLs or spending 15 minutes on a site. We determined these termination limits by running test crawls on random samples of shopping websites. Finally, we opted to extract no more than five product pages from each shopping website.

To evaluate our crawler's performance, we randomly sampled 100 shopping websites from our corpus of 19K shopping websites and examined the product URLs the crawler returned for each of these websites. For 86 of those 100 websites, our crawler successfully extracted and returned legitimate product pages where they were present, and it returned no product pages where there were not any. For the remaining 14 websites, the crawler either timed out because the website was no longer reachable, the website included a step that the crawler could not handle (e.g., the website required selecting a country of origin), or the `Add to cart' button was incorrectly detected. We then used the crawler on all of the 19K shopping websites, and in total we gathered 53,180 product pages from 11,286 shopping websites.

\subsubsection{Simulating Product Purchase Flows}
To simulate a user's typical shopping flow---which included selecting certain product options (e.g., size or color), adding the product to the cart, viewing the cart, and checking out---we designed and built an interactive `checkout crawler'.

We based our checkout crawler on OpenWPM, a fully instrumented browser platform that is designed to conduct large-scale privacy and web-tracking measurement studies~\cite{englehardt-tracking-2017}. We extended OpenWPM in a number of ways to interact with the product pages we collected previously, including identifying various interface elements using scoring functions similar to the ones we described in Section~\ref{section:collection}. Each of these functions would output the most likely `Add to cart' buttons, `View cart' buttons, and `Checkout' buttons, which the crawler would click in--order across multiple pages. Because websites do not follow uniform HTML markup and design, our crawler needed to account for a variety of design alternatives and edge cases to simulate user interaction, such as dismissing popup dialogs, and identifying and interacting with product options (e.g., selecting a size and color for a t-shirt) to add a product to cart.

We collected three types of data during this crawl for each product page. First, we saved the page source on visit. Second, we took screenshots each time the state of the page changed (e.g., clicking a button or selecting a product option). Third, we extended OpenWPM's HTTP instrumentation to store HTTP Archive (HAR)~\cite{har-wiki}) files for each crawled page since HAR files are not limited to HTTP headers and contain full response contents that can be used for further analysis.

To evaluate our crawler's performance, we randomly sampled 100 product pages from the crawl in Section~\ref{section:producturl} and examined whether our crawler was able to simulate a user's shopping flow. In 66 of the 100 pages, our crawler reached the checkout page successfully. In 14 of the remaining 34, the crawler was able to add the product to cart but it was unable to proceed to the cart page; most often this was the result of complex product interaction (e.g., selecting the dimensions of a rug), which our crawler was not designed to perform. In the remaining 20 cases, either we produced Selenium exceptions, or failed to discover cart and checkout buttons. We then used the crawler on all of the 53K product pages. We divided the 53K product URLs into two equal-length lists to reduce the total crawling time. These crawls took approximately 90 hours to complete.

\subsubsection{Capturing Meaningful Text Using Page Segmentation}
The checkout crawler divided all the pages it visited into meaningful page segments to help discover dark patterns. These segments can be thought of as `building blocks' of web pages, representing meaningful smaller sections of a web page. These formed the basic units for our data analysis and clustering.

We defined segments as \emph{visible} HTML elements that contained no other block-level elements~\cite{block-level-MDN} and contained at least one text element---that is, elements of type ~\texttt{TEXT\_NODE}~\cite{node-types-MDN}. However, since websites may use a virtually endless variety of markup and designs, we iteratively developed our segmentation algorithm, testing it on samples of shopping websites and accounting for possible edge cases. Algorithm~\ref{alg:segments} and Figure~\ref{fig:segmentation} in the Appendix detail the segmentation algorithm and illustrate its output for one web page, respectively.

Before segmenting each web page, the crawler waited for the page to load completely, also accounting for the time needed for popup dialogs to appear. However, web pages may also display text from subsequent user interactions, and with dynamically loaded content (e.g., a countdown timer). To capture possible segments from such updates to the web page during a crawl---no matter how minor or transient---we integrated the Mutation Summary \cite{mutation-summary-github} library into our checkout crawler. The Mutation Summary library combines DOM \texttt{MutationObserver} events \cite{mutation-observer-MDN} into compound event summaries that are easy to process. When the checkout crawler received a new Mutation Summary representing updates to the page, it segmented (Algorithm~\ref{alg:segments}) this summary and stored the resulting segments. 

For each segment, we stored its HTML Element type, its element text (via \texttt{innerText}), its dimensions and coordinates on the page, and its style including its text and background colors. Our crawls resulted in $\sim$13 million segments across the 53K product URL pages.

\subsection{Data Analysis with Clustering}
We employed hierarchical clustering to discover dark patterns from the data set of segments. Our use of clustering was not to discover a set of latent constructs in the data but rather to organize the segments in a manner that would be conducive to scanning, making it easier for an expert analyst to sift through the clusters for possible dark patterns.

\subsubsection{Data Preprocessing}

Many of the $\sim$13 million segments collected during our crawls were duplicates, such as multiple `Add to cart' segments across multiple websites. Since we only used text-based features for our analyses, we retained unique pieces of text across the websites in our data set (e.g., one segment containing the text `Add to cart' across all the websites in our data set). We also replaced all numbers with a placeholder before performing this process to further reduce duplicates. This preprocessing reduced the set of segments by 90\% to $\sim$1.3 million segments.

\subsubsection{Feature Representations and Hierarchical Clustering}

Before performing clustering, we transformed the text segments into a Bag of Words (BoW) representation. Each entry in the resulting BoW matrix ($M_{ij}$) indicated the number of times token $j$ appeared in segment $i$.\footnote{We did not use the Term Frequency-Inverse Document Frequency (TF-IDF) representation as upon clustering, it resulted in anywhere between 70\%-75\% of the segments being classified as noise. We believe this may have been because of the incorrect IDF scaling factor since the segments were not all drawn from a pool of independent observations---i.e., multiple segments originated from the same website} We filtered all stop words\footnote{Using Python \texttt{NLTK}~\cite{nltk}} and punctuation---except currency symbols, since these are indicative of product price---from the list of tokens, and further only retained tokens that appeared in at least 100 segments. This resulted in a vocabulary of 10,133 tokens.

Given this large size of our vocabulary---and thus the dimensions of the segment-token matrix---we performed Principal Component Analysis (PCA) on the BoW matrix. We retained 3 components from the PCA, which together captured more than 95\% of the variance in the data.

We used the Hierarchical Density-Based Spatial Clustering of Applications with Noise (HDBSCAN) algorithm~\cite{hdbscan-paper} implemented in the \texttt{HDBSCAN} Python library~\cite{hdbscan-library} to extract clusters from this data. We chose HDBSCAN over other clustering algorithms since it is robust to noise in the data, and it allows us to vary the minimum size of the clusters (\texttt{min\_cluster\_size}). We varied a total of four passes at clustering: two ~\texttt{min\_cluster\_size} values (5 and 10) $\times$ two distance metrics (Manhattan distance or L1 norm, and Euclidean distance or L2 norm). We picked sufficiently small values for the \texttt{min\_cluster\_size} parameter to keep the size of the noise cluster small and to avoid coercing segments into one cluster.

The clustering output across the BoW input was nearly the same. As expected, a \texttt{min\_cluster\_size} of 10 resulted in a larger noise cluster compared to a \texttt{min\_cluster\_size} of 5---but only marginally larger regardless of the distance metric. However, since the \texttt{min\_cluster\_size} of 10 produced significantly fewer clusters, we picked its output over the others. It contained 10,277 clusters.

\subsubsection{Examining and Analyzing the Clusters}

Once the clustering was complete, we made two passes through the data. The goal of pass one was to include clusters that contained any segments that might manifest as dark patterns. In this pass, one researcher scanned the clusters and identified possible clusters of interest, recording all those clusters that represented specific types of user interfaces (e.g., login choices, cart totals), website characteristics (e.g., stock notifications), and product options (e.g., small/medium/large) that generally appear on shopping websites. This step filtered down the clusters from 10,277 to 1,768.

In pass two, we extracted all the websites that corresponded to these segments for further examination. The research team used the literature on dark patterns~\cite{brignull-dark-patterns,gray-dark-patterns-2018,nodder2013evil} and impulse buying~\cite{moser2019impulse}, and media coverage of high-pressure sales and marketing tactics (e.g., ~\cite{hpst}) to create a shared understanding of possible dark patterns using the examples cited in these works to guide our thinking. In order to validate the coding of clusters, two researchers examined a sample of 200 of the 1,768 clusters, and recorded any dark patterns they encountered. The researchers also examined each website's set of screenshots and visited the websites to gain context and additional information surrounding the segments (e.g., discovering practices associated with the flagged pattern). To measure agreement between the researchers, we computed Cohen's kappa between the segments that were recorded---resulting in a score of 0.74. The team discussed and resolved all disagreements, and one researcher then examined the remaining clusters in the same manner. The team then discussed the resulting dark patterns, and iteratively grouped them into types and broader categories.

\subsection{Detecting Deceptive Dark Patterns}
\label{section:detecting-deceptive-adp}

We further examined many of the dynamic dark patterns---those patterns that displayed transient values (e.g., a countdown timer)---for deceptive practices. To this end, we used our checkout crawler to `monitor' the websites containing dark patterns of interest once every four hours for a period of five days. We combined this data with several dark pattern-specific heuristics---which we describe in the following sections---to uncover instances of deceptive practices.


\section{Findings} 
\label{section:findings}

\begin{figure}
    \centering
    \includegraphics[width=0.6\columnwidth]{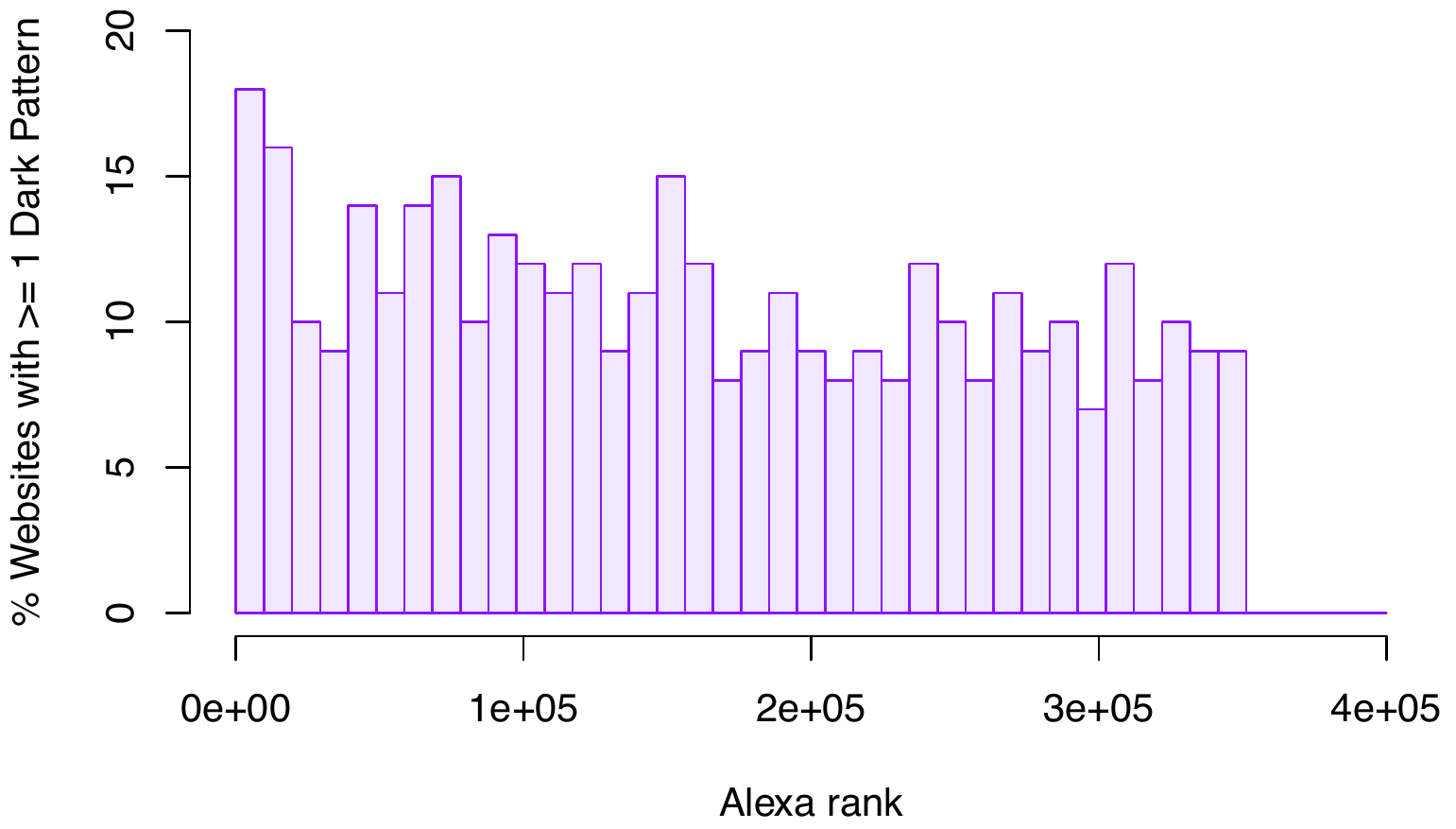}
    \caption{Distribution of the dark patterns we discovered over the Alexa rank of the websites. Each bin indicates the percentage of shopping websites in that bin that contained at least one dark pattern.}
    \label{fig:distribution}
\end{figure}

In total, we discovered \totaldpinst{} instances of dark patterns from \totaldpweb{}  ($\sim$\totaldpperc{}) websites in our data set of 11K shopping websites. Given that \begin{enumerate*} \item our crawler only explored the product pages, cart pages, and checkout pages of websites, \item our analyses only took text-based user interfaces into account, \end{enumerate*} this number represents a lower-bound estimate of the prevalence of dark patterns. We divide our discussion of the findings by first illustrating the categories of dark patterns revealed by our analyses, and then by describing our findings on the ecosystem of third-parties that enable dark patterns.

\def\arraystretch{1.5}
\begin{table}[]
\scriptsize
\caption{Categories and types of dark patterns along with their description, prevalence, and definitions. \newline Legend: \CIRCLE = Always, \LEFTcircle = Sometimes, \Circle = Never}

\begin{tabular}{p{1.2cm}lp{4.25cm}|ll|p{0.09cm}p{0.09cm}p{0.09cm}p{0.09cm}p{0.09cm}|p{1cm}}
\multicolumn{1}{c}{\textbf{Category}} & \multicolumn{1}{c}{\textbf{Type}} & \multicolumn{1}{c|}{\textbf{Description}} & \rot{\textbf{\# Instances}} & \rot{\textbf{\# Websites}} & \rot{\textbf{Asymmetric?}} & \rot{\textbf{Covert?}} & \rot{\textbf{Deceptive?}} & \rot{\textbf{Hides Info?}} & \rot{\textbf{Restrictive?}} & \multicolumn{1}{c}{\textbf{\begin{tabular}[c]{@{}c@{}}Cognitive\\ Biases\end{tabular}}} \\ \hline

Sneaking & Sneak into Basket & Adding additional products to users' shopping carts without their consent & \sibi{} & \sibw{} & \Circle & \Circle & \LEFTcircle & \CIRCLE & \Circle & Default Effect \\

 & Hidden Costs & Revealing previously undisclosed charges to users right before they make a purchase & \hci{} & \hcw{} & \Circle & \Circle & \LEFTcircle & \CIRCLE & \Circle & Sunk Cost Fallacy \\
 
 
 & Hidden Subscription & Charging users a recurring fee under the pretense of a one-time fee or a free trial & \hsi{} & \hsw{} & \Circle & \Circle & \LEFTcircle & \CIRCLE & \Circle & None \\ \hline
 
Urgency & Countdown Timer & Indicating to users that a deal or discount will expire using a counting-down timer & \cti{} & \ctw{} & \Circle & \LEFTcircle & \LEFTcircle & \Circle & \Circle & Scarcity Bias \\

 & Limited-time Message & Indicating to users that a deal or sale will expire will expire soon without specifying a deadline & \lti{} & \ltw{} & \Circle & \LEFTcircle & \Circle & \CIRCLE & \Circle & Scarcity Bias \\ \hline

Misdirection & Confirmshaming & Using language and emotion (shame) to steer users away from making a certain choice & \csi{} & \csw{} & \CIRCLE & \Circle & \Circle & \Circle & \Circle & Framing Effect \\

 & Visual Interference & Using style and visual presentation to steer users to or away from certain choices & \vii{} & \viw{} & \LEFTcircle & \CIRCLE & \LEFTcircle & \Circle & \Circle & Anchoring \& Framing Effect \\
 
 & Trick Questions & Using confusing language to steer users into making certain choices & \tqi{} & \tqw{} & \CIRCLE & \CIRCLE & \Circle & \Circle & \Circle & Default \& Framing Effect \\
 
& Pressured Selling & Pre-selecting more expensive variations of a product, or pressuring the user to accept the more expensive variations of a product and related products & \psei{} & \psew{} & \LEFTcircle & \LEFTcircle & \Circle & \Circle & \Circle & Anchoring \& Default Effect, Scarcity Bias \\ \hline

Social Proof & Activity Message & Informing the user about the activity on the website (e.g., purchases, views, visits) & \ani{} & \anw{} & \Circle & \LEFTcircle & \LEFTcircle & \Circle & \Circle & Bandwagon Effect \\

 & Testimonials & Testimonials on a product page whose origin is unclear & \testi{} & \testw{} & \Circle & \Circle & \LEFTcircle & \Circle & \Circle & Bandwagon Effect \\ \hline
 
Scarcity & Low-stock Message & Indicating to users that limited quantities of a product are available, increasing its desirability & \lsi{} & \lsw{} & \Circle & \LEFTcircle & \LEFTcircle & \LEFTcircle & \Circle & Scarcity Bias \\

 & High-demand Message & Indicating to users that a product is in high-demand and likely to sell out soon, increasing its desirability & \hdi{} & \hdw{} & \Circle & \LEFTcircle & \Circle & \Circle & \Circle & Scarcity Bias \\ \hline
 
Obstruction & Hard to Cancel & Making it easy for the user to sign up for a service but hard to cancel it & \htci{} & \htcw{} & \Circle & \Circle & \Circle & \LEFTcircle & \CIRCLE & None \\ \hline

Forced Action & Forced Enrollment & Coercing users to create accounts or share their information to complete their tasks  & \fai{} & \faw{} & \CIRCLE & \Circle & \Circle & \Circle & \CIRCLE & None \\ \hline

\end{tabular}
\label{tab:types}
\end{table}

\subsection{Categories of Dark Patterns}

Our analyses revealed 15 types of dark patterns contained in 7 broader categories. Where applicable, we use the dark pattern labels proposed by Gray et al.\ ~\cite{gray-dark-patterns-2018} and Brignull~\cite{brignull-dark-patterns} to describe these types and categories. Table~\ref{tab:types} summarizes our findings, highlighting the number of separate instances of dark patterns found for each type. 

Figure~\ref{fig:distribution} shows the distribution of the websites containing dark patterns over their Alexa ranks. The distribution suggests that dark patterns are more likely to appear on popular websites (Spearman's Rho = -0.62, $p < 0.0001$). In the following sections, we describe the various categories and types of dark patterns we discovered.

\subsubsection{Sneaking}

\begin{figure}[t]
\centering
\subfloat[Sneak into Basket on \texttt{avasflowers.net}. Despite requesting no greeting cards, one worth \$3.99 is automatically added.\label{fig:sneaking:sb}]{%
{%
\setlength{\fboxsep}{1pt}%
\setlength{\fboxrule}{0.4pt}%
\fbox{\includegraphics[width=\textwidth]{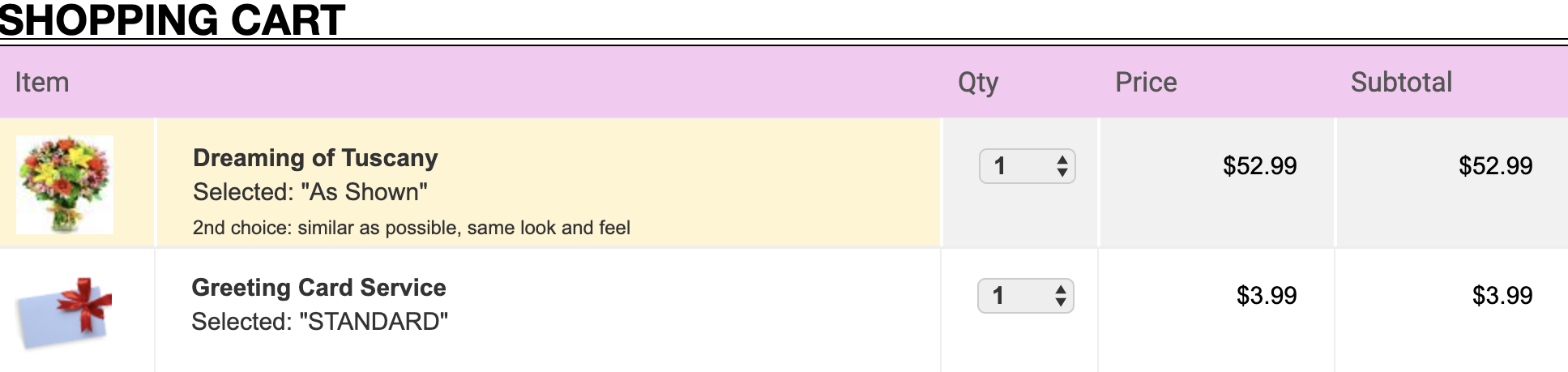}}}
}

\subfloat[Hidden Costs on \texttt{proflowers .com}. The Care \& Handling charge (\$2.99) is disclosed on the last step.\label{fig:sneaking:hc}]{%
{%
\setlength{\fboxsep}{3pt}%
\setlength{\fboxrule}{0.4pt}%
\fbox{\includegraphics[height=3cm]{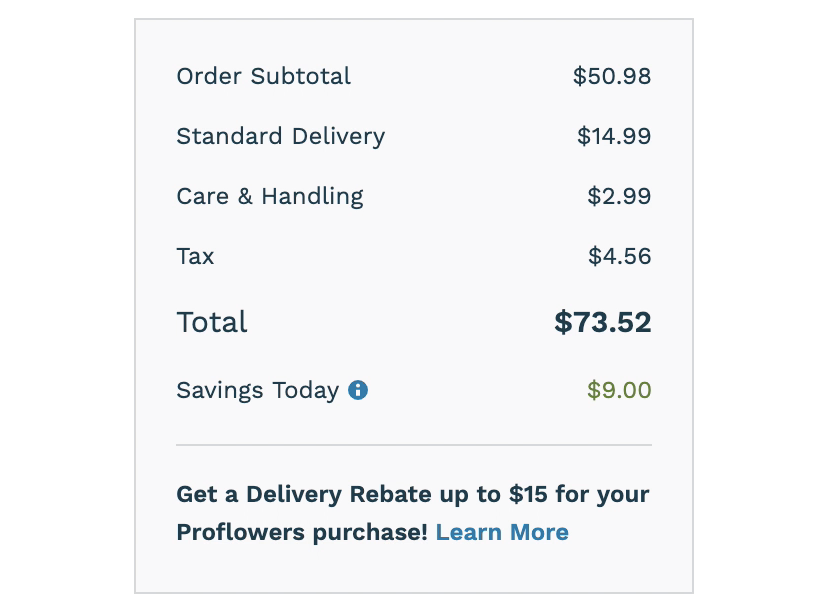}}%
}%
}
\hspace*{\fill}
\subfloat[Hidden Subscription on \texttt{wsjwine.com}. Left: The website fails to disclose that the \emph{Advantage} service is an annual subscription worth \$89 unless the user clicks on \emph{Learn More}. Right: The service in cart. \label{fig:sneaking:hs}]{%
{%
\setlength{\fboxsep}{3pt}%
\setlength{\fboxrule}{0.4pt}%
\fbox{\includegraphics[height=3cm]{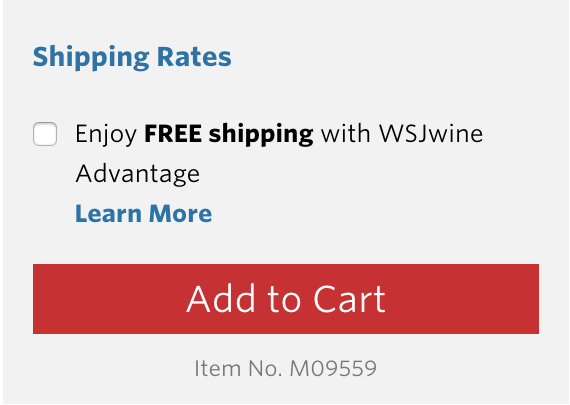}
\includegraphics[height=3cm]{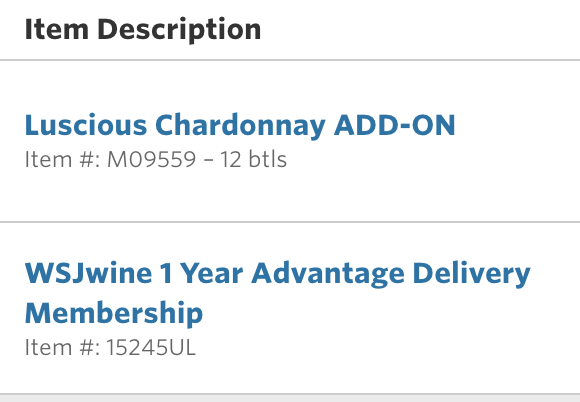}}
}
}

\caption{Three types of the Sneaking category of dark patterns.}
	\label{fig:sneaking}
\end{figure}

Coined by Gray et al.\ in their taxonomy~\cite{gray-dark-patterns-2018}, `Sneaking' refers to the category of dark patterns that attempt to misrepresent user actions, or hide/delay information that, if made available to users, they would likely object to. We observed three types of the Sneaking dark pattern: Sneak into Basket~\cite{brignull-dark-patterns}, Hidden Costs~\cite{brignull-dark-patterns}, and Hidden Subscription (Brignull's Forced Continuity~\cite{brignull-dark-patterns}) on \sneakw{} shopping websites. Figure~\ref{fig:sneaking} highlights instances of these three types.

\paragraph{\textbf{Sneak into Basket}} The `Sneak into Basket' dark pattern adds additional products to users' shopping carts without their consent, often promoting the added products as `bonuses' and `necessary'. Sneak into Basket exploits the default effect cognitive bias in users, with the website behind it hoping that users will stick with the products it adds to cart. One instance of Sneak into Basket is shown in Figure~\ref{fig:sneaking:sb}, where adding a bouquet of flowers to the shopping cart on \texttt{avasflowers.net} also adds a greeting card. In another instance on \texttt{laptopoutlet.co.uk} ---not shown in the figure--- adding an electronic product, such as a laptop, to the shopping cart also adds product insurance. Other websites, such as \texttt{cellularoutfitter.com}, add additional products (e.g., a USB charger) to the shopping cart using pre-selected checkboxes. While such checkboxes could be deselected by a vigilant user, the additional products would be added by default in the absence of any intervention. In our data set, we found a total of \sibi{} instances of the Sneak into Basket dark pattern.

Using our taxonomy of dark pattern characteristics, we classify Sneak into Basket as at least partially \emph{deceptive} (it incorrectly represents the nature of the action of adding an item to the shopping cart) and \emph{information hiding} (it deliberately disguises how the additional products were added to cart from users) in nature. However, it is not \emph{covert}: users can visibly see and realize that the website included additional products to their shopping carts.

\paragraph{\textbf{Hidden Costs}}\label{section:hidden-costs} The `Hidden Costs' dark pattern reveals new, additional, and often unusually high charges to users just before they are about to complete a purchase. Examples of such charges include `service fees' or `handling costs'. Often these charges are only revealed at the end of a checkout process, after the user has already filled out shipping/billing information, and consented to terms of use. The Hidden Costs dark pattern exploits the sunk cost fallacy cognitive bias: users are likely to feel so invested in the process that they justify the additional charges by completing the purchase to not waste their effort. Figure~\ref{fig:sneaking:hc} shows the Hidden Costs dark pattern on \texttt{proflowers.com}, where the `Care \& Handling' charge of \$2.99 is revealed immediately before confirming the order. In our data set, we found a total of \hci{} instances of the Hidden Costs dark pattern.

Using our taxonomy of dark pattern characteristics, we classify Hidden Costs as at least partially \emph{deceptive} (it relies on minimizing and delaying information from users), and thus also \emph{information hiding} in nature. Like Sneak into Basket, Hidden Costs is not \emph{covert}: users can visibly see and realize that the website included additional charges.

\paragraph{\textbf{Hidden Subscription}} The `Hidden Subscription' dark pattern charges users a recurring fee under the pretense of a one-time fee or a free trial. Often, if at all, users become aware of the recurring fee once they are charged several days or months after their purchase. For instance, we discovered that \texttt{wsjwine.com} offers users an \emph{Advantage} service which appears to be a one-time payment of \$89 but renews annually, as shown in Figure~\ref{fig:sneaking:hs}. Further, Hidden Subscription often appears with the `Hard to Cancel' dark pattern---which we describe in Section \ref{section:obstruction}---thereby making the recurring charges harder to cancel than signing up for them. In our data set, we found a total of \hsi{} instances of Hidden Subscription dark pattern.

Using our taxonomy of dark pattern characteristics, we classify Hidden Subscription as at least partially \emph{deceptive} (it misleads users about the nature of the initial offer) and \emph{information hiding} (it withholds information about the recurring fees from users) in nature.

\subsubsection{Urgency}
\label{section:urgency}

`Urgency' refers to the category of dark patterns that impose a deadline on a sale or deal, thereby accelerating user decision-making and purchases~\cite{time-aggarwal, time-inman, cialdini2009influence,nodder2013evil}. Urgency dark patterns exploit the scarcity bias in users---making discounts and offers more desirable than they would otherwise be, and signaling that inaction would result in losing out on potential savings. These dark patterns create a potent `fear of missing out' effect particularly when combined with the Social Proof (Section~\ref{section:findings:social-proof}) and Scarcity (Section~\ref{section:scarcity}) dark patterns.

\begin{figure}[t]
\subfloat[Countdown Timer on \texttt{mattressfirm.com}. The header displays a \emph{Flash Sale} where the majority of discounted products remain the same on a day-to-day basis.\label{fig:urgency:ct1}]{%
{%
\setlength{\fboxsep}{3pt}%
\setlength{\fboxrule}{0.4pt}%
\fbox{\includegraphics[height=2cm]{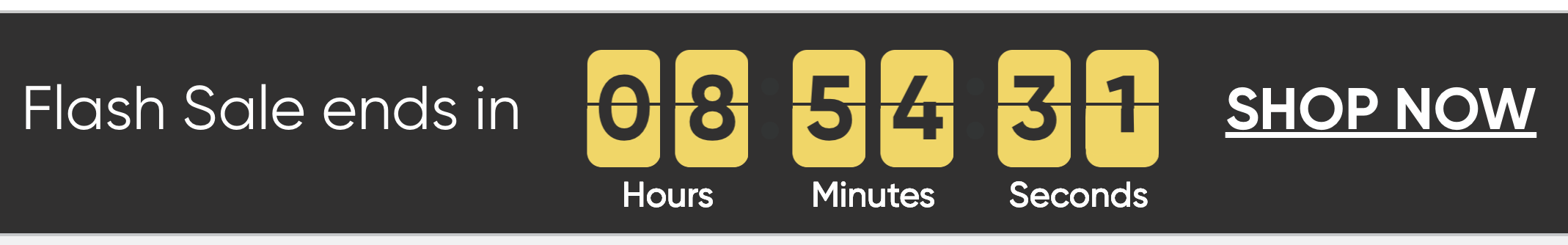}}}
}

\subfloat[Countdown Timer on \texttt{justfab.com}. The offer is available even after the timer expires.\label{fig:urgency:ct2}]{%
{%
\setlength{\fboxsep}{3pt}%
\setlength{\fboxrule}{0.4pt}%
\fbox{\includegraphics[height=2.9cm]{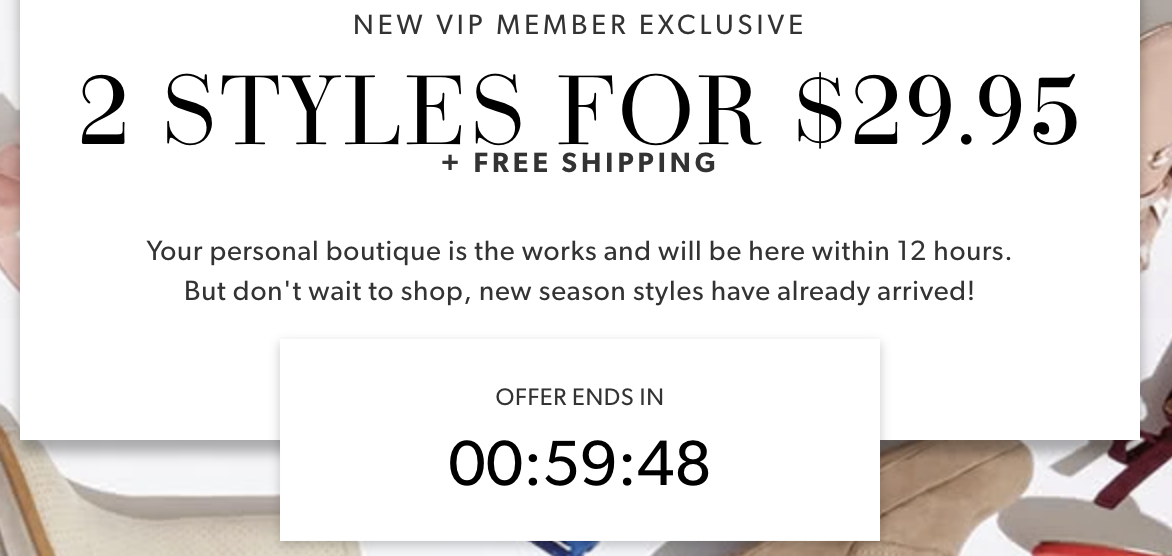}}%
}%
}
\hspace*{\fill}
\subfloat[Limited-time Message on \texttt{chicwish.com}. The website claims the sale will end `soon' without stating a deadline. \label{fig:urgency:lt}]{%
{%
\setlength{\fboxsep}{3pt}%
\setlength{\fboxrule}{0.4pt}%
\fbox{\includegraphics[height=2.9cm]{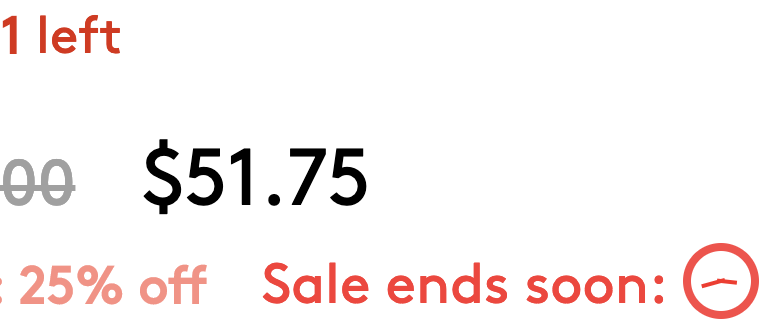}}
}
}

\caption{Two types of the Urgency category of dark patterns.}
	\label{fig:urgency}
\end{figure}

We observed two types of the Urgency dark pattern: Countdown Timers and Limited-time Messages on \urgw{} shopping websites across their product, cart, and checkout pages. In product pages, these indicated deadlines about site-wide sales and coupons, sales on specific products, or shipping deadlines; in cart pages, they indicated deadlines about product reservation (e.g., `Your cart will expire in 10:00 minutes, please check out now') and coupons, urging users to complete their purchase. Figure~\ref{fig:urgency} highlights instances of these two types. 



\paragraph{\textbf{Countdown Timers}} The `Countdown Timer' dark pattern is a dynamic indicator of a deadline, counting down until the deadline expires. Figures~\ref{fig:urgency:ct1} and~\ref{fig:urgency:ct2} show the Countdown Timer dark pattern on \texttt{mattressfirm.com} and \texttt{justfab.com}, respectively. One indicates the deadline for a recurring \emph{Flash Sale}, the other a \emph{Member Exclusive}. In our data set, we found a total of \cti{} instances of the Countdown Timer dark pattern.


\paragraph{Deceptive Countdown Timers.} Using the visit-and-record method described in Section~\ref{section:detecting-deceptive-adp}, we examined the countdown timers in our data set for deceptive practices. We stitched the screenshots of each countdown timer from the repeated visits of our crawler to a website into a video, and viewed the resulting videos to observe the behavior of the timers. We considered a countdown timer deceptive if \begin{enumerate*} \item the timer reset after timeout with the same offer still valid, or \item the timer expired but the offer it claimed was expiring was still valid even following expiration.\end{enumerate*}

In our data set, we discovered a total of \ctid{} instances of deceptive Countdown Timers on \ctwd{} shopping websites. One such example is shown in Figure~\ref{fig:urgency:ct2} on \texttt{justfab.com}, where the advertised offer remains valid even after the countdown timer of 60 minutes expires.

Using our taxonomy of dark pattern characteristics, we classify Countdown Timers as partially \emph{covert} (it creates a heightened sense of immediacy, unbeknownst to at least some users), and sometimes \emph{deceptive} (it can mislead users into believing an offer is expiring when in reality it is not) in nature.

\paragraph{\textbf{Limited-time Messages}} Unlike Countdown Timers, the `Limited-time Message' dark pattern is a static urgency message without an accompanying deadline. By not stating the deadline, websites withhold information from users, and thus misrepresent the nature of the offer~\cite{asa}. Figure ~\ref{fig:urgency:lt} shows an instance of the Limited-time Message dark pattern on \texttt{chicwish.com}, where the advertised sale is stated to end `soon' with no mention of the end date. For every such instance we discovered, we verified that the shopping website made no disclosure about the accompanying deadline (e.g., in the fine print and in the terms of sale pages). In our data set, we discovered a total of \lti{} instances of the Limited-time Message dark pattern.

Using our taxonomy of dark pattern characteristics, we classify Limited-time Messages as at least partially \emph{covert} (similar to Countdown Timers), and \emph{information hiding} (unlike Countdown Timers, they do not reveal the deadline in their offers) in nature.

\subsubsection{Misdirection}

The `Misdirection' category of dark patterns uses visuals, language, and emotion to steer users toward or away from making a particular choice. Misdirection functions by exploiting different affective mechanisms and cognitive biases in users without actually restricting the set of choices available to users. Our version of the Misdirection dark pattern is inspired by Brignull's original Misdirection dark pattern~\cite{brignull-dark-patterns}. However, while Brignull considered Misdirection to occur exclusively using stylistic and visual manipulation, we take a broader view of the term, also including Misdirection caused by language and emotional manipulation.

\begin{figure}[t]
\centering
\subfloat[Confirmshaming on \texttt{radioshack.com}. The option to dismiss the popup is framed to shame the user into avoiding it. \label{fig:misdirection:confirmshaming}]{
{
\setlength{\fboxsep}{3pt}%
\setlength{\fboxrule}{0.4pt}%
\fbox{
\includegraphics[height=2.6cm]{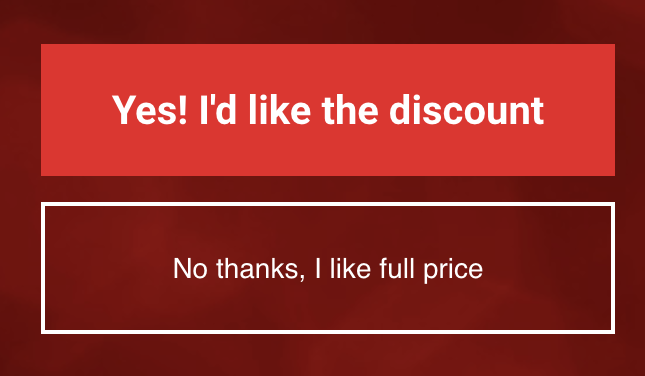}
}
}
}
\hspace*{\fill}
\subfloat[Visual Interference on \texttt{greenfingers.com}. The option to opt out of marketing communication is grayed, making it seem unavailable even though it can be clicked. \label{fig:misdirection:visual}]{%
{%
\setlength{\fboxsep}{3pt}%
\setlength{\fboxrule}{0.4pt}%
\fbox{\includegraphics[height=2.6cm]{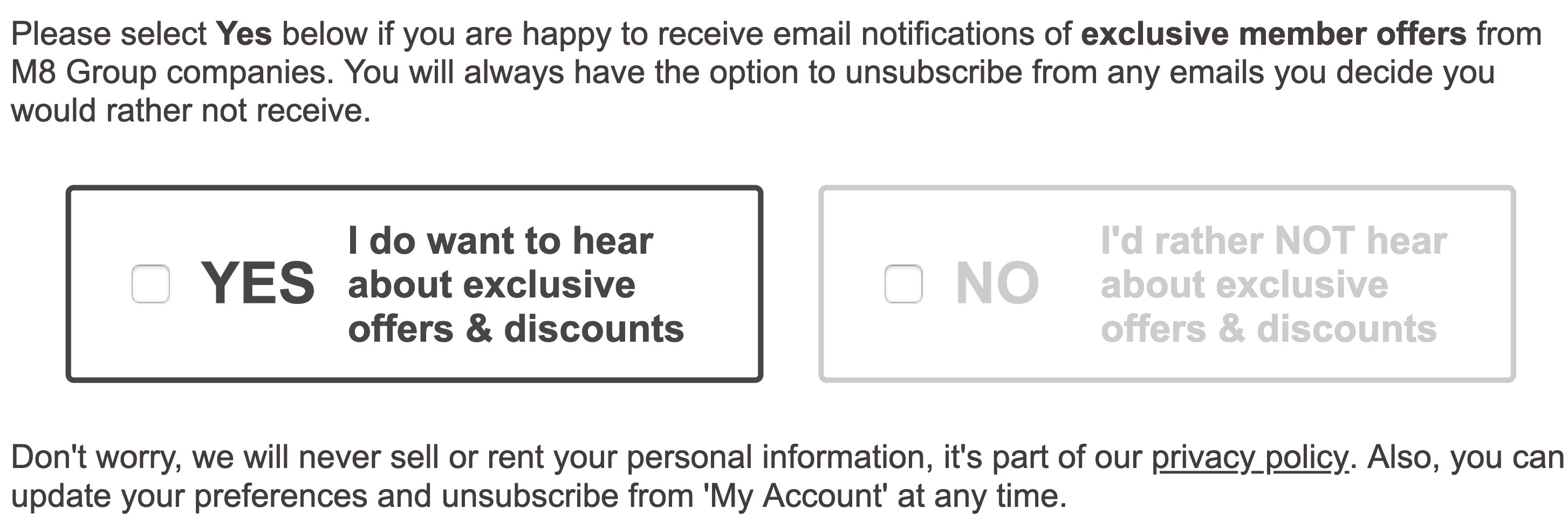}}
}
}

\subfloat[Trick Questions on \texttt{newbalance.co.uk}. Opting out of marketing communication requires ticking the checkbox. \label{fig:misdirection:trickquestions}]{
{
\setlength{\fboxsep}{1pt}%
\setlength{\fboxrule}{0.4pt}%
\fbox{
\includegraphics[height=1.35cm]{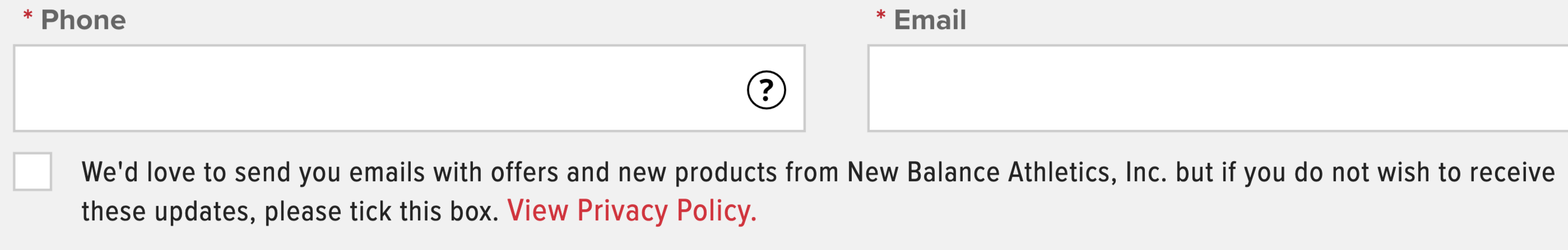}
}
}
}
\hspace*{\fill}
\subfloat[Pressured Selling on \texttt{1800flowers.com}. The most expensive product is the default.\label{fig:misdirection:ps}]{%
{%
\setlength{\fboxsep}{1pt}%
\setlength{\fboxrule}{0.4pt}%
\fbox{\includegraphics[height=1.35cm]{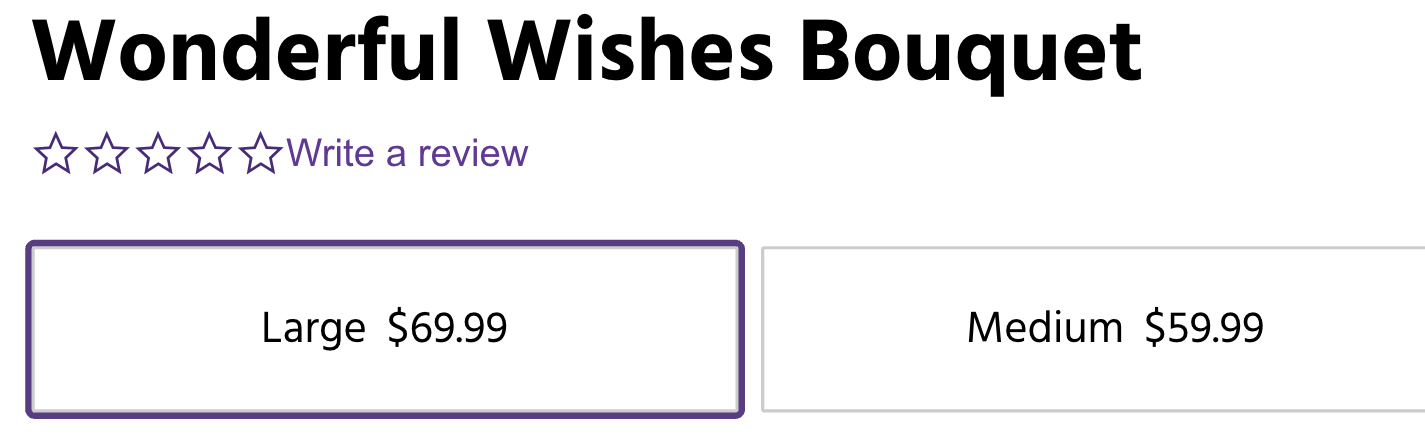}}
}
}

\caption{Four types of the Misdirection category of dark patterns.}
	\label{fig:misdirection}
\end{figure}

We observed four types of the Misdirection dark pattern: Confirmshaming~\cite{brignull-dark-patterns}, Trick Questions~\cite{brignull-dark-patterns}, Visual Interference~\cite{gray-dark-patterns-2018}, and Pressured Selling on \mdw{} shopping websites. Figure~\ref{fig:misdirection} highlights instances of these four types.

\paragraph{\textbf{Confirmshaming}} Coined by Brignull \cite{brignull-dark-patterns}, the `Confirmshaming' dark pattern uses language and emotion to steer users away from making a certain choice. Confirmshaming appeared most often in popup dialogs that solicited users' email addresses in exchange for a discount, where the option to decline the offer---which the website did not want users to select---was framed as a shameful choice. Examples of such framing included `No thanks, I like paying full price', `No thanks, I hate saving money', and `No thanks, I hate fun \& games'. By framing the negative option as such, the Confirmshaming dark pattern exploits the framing effect cognitive bias in users and shame, a powerful behavior change agent~\cite{lickel2014shame}. Figure~\ref{fig:misdirection:confirmshaming} shows one instance of the Confirmshaming dark pattern on \texttt{radioshack.com}. In our data set, we found a total of \csi{} such instances.

Using our taxonomy of dark pattern characteristics, we classify Confirmshaming as \emph{asymmetric} (the opt-out choice shames users into avoiding it) in nature. However, Confirmshaming is not \emph{covert}, since users can visibly see and realize that the design is attempting to influence their choice.

\paragraph{\textbf{Visual Interference}} The `Visual Interference' dark pattern uses style and visual presentation to influence users into making certain choices over others (Brignull's original description of Misdirection~\cite{brignull-dark-patterns}). Although we excluded style information in our clustering analysis, we extracted these patterns as a consequence of examining the text the patterns displayed. In some instances, websites used the Visual Interference dark pattern to make certain courses of action more prominent over others. For example, the subscription offering on \texttt{exposedskincare.com} is stylistically more prominent and emphasized than the non-subscription offering. In other instances, websites used visual effects on textual descriptions to inflate the discounts available for products. For example, websites such as \texttt{dyson.co.uk} and \texttt{justfab.com} offered free gifts to users, and then used these gifts to inflate the savings on users' purchases in the checkout page---even when the originally selected product was not on discount. In one instance on \texttt{greenfingers.com}, we discovered that the option to decline marketing communication is greyed out, creating an illusion that the option is unavailable or disabled even though it can be clicked, as shown in Figure~\ref{fig:misdirection:visual}. In our data set, we found a total of \vii{} instances of the Visual Interference dark pattern.

Using our taxonomy of dark pattern characteristics, we classify Visual Interference as sometimes \emph{asymmetric} (in some instances it creates unequal choices, steering users into one choice over the other), \emph{covert} (users may not realize the effect the visual presentation has had on their choice), and sometimes \emph{deceptive} (e.g., when a website presents users with a `lucky draw' from a list of potential deals, but the draw process is deterministic unbeknownst to users) in nature.

\paragraph{\textbf{Trick Questions}} Also originating from Brignull's taxonomy \cite{brignull-dark-patterns}, the `Trick Questions' dark pattern uses confusing language to steer users into making certain choices. Like Confirmshaming, Trick Questions attempt to overcome users' propensity to opt out of marketing and promotional messages by subtly inverting the entire opt-out process. Most often, websites achieved this effect by introducing confusing double negatives (e.g., `Uncheck the box if you prefer not to receive email updates'), or by using negatives to alter expected courses of action, such as checking a box to opt out (e.g., `We would like to send you emails. If you do not wish to be contacted via email, please ensure that the box is not checked'). 

We note here that we only considered an opt-out choice as a Trick Question dark pattern when it was misleading, such as when the user has to check a box and the text began with an affirmative statement about the undesirable practice (e.g., `We want to send you marketing email...') since these would more likely be missed by users as opposed to ones that began with the opt-out choice (e.g., `Please tick here to opt-out of...').\footnote{We note that while Gray et al. ~\cite{gray-dark-patterns-2018} consider the latter as Trick Questions, we do not take that stance. However, we do consider all opt-out messages as concerning. We discovered 23 instances of opt-out choices that did not begin with an affirmative statement in total.} Trick Questions exploits the default and framing effect cognitive biases in users, who become more susceptible to a choice they erroneously believe is aligned with their preferences. Figure~\ref{fig:misdirection:trickquestions} shows one instance of Trick Questions on \texttt{newbalance.co.uk}. In our data set, we found a total of \tqi{} such instances, occurring most often during the checkout process when collecting user information to complete purchases.

Using our taxonomy of dark pattern characteristics, we classify Trick Questions as \emph{asymmetric} (opting out is more burdensome than opting in) and \emph{covert} (users fail to understand the effect of their choice as a consequence of the confusing language) in nature.

\paragraph{\textbf{Pressured Selling}} The `Pressured Selling' dark pattern refers to defaults or often high-pressure tactics that steer users into purchasing a more expensive version of a product (\textit{upselling}) or into purchasing related products (\textit{cross-selling}). The Pressured Selling dark pattern exploits a variety of different cognitive biases, such as the default effect, the anchoring effect, and the scarcity bias to drive user purchasing behavior. Figure~\ref{fig:misdirection:ps} shows one such instance on \texttt{1800flowers.com}, where the largest flower bouquet is selected by default. The dark pattern makes the most expensive option the point of comparison---an `anchor'---and thus increases the probability of users overlooking the least expensive option~\cite{parkchoosing}. In another instance, on \texttt{fashionworld.co.uk}, the website opened popup dialogs that the user had to explicitly decline immediately after adding a product to cart. These dialogs urged users to buy more `Hot sellers', `Deals', and `Bundled' products. In our data set, we found a total of \psei{} instances of the Pressured Selling dark pattern.

Using our taxonomy of dark pattern characteristics, we classify Pressured Selling as sometimes \emph{asymmetric} (it pushes users towards accepting more expensive product options) and at least partially \emph{covert} (users fail to realize that they have purchased a more expensive product than they would have, had they been defaulted with the least expensive product to begin with) in nature.

\subsubsection{Social Proof}
\label{section:findings:social-proof}

According to the social proof principle, individuals determine the correct action and behavior for themselves in a given situation by examining the action and behavior of others~\cite{cialdini2009influence,nodder2013evil}. The `Social Proof' dark pattern uses this influence to accelerate user decision-making and purchases, exploiting the bandwagon effect cognitive bias to its advantage. Studies have shown that individuals are more likely to impulse buy when shopping with their peers and families~\cite{luo-shopping-2005}. 

\begin{figure}[t]
\centering
\subfloat[Activity Notification on \texttt{tkmaxx.com}. The message indicates how many people added the product to the cart in the last 72 hours.\label{fig:sp:an1}]{
{
\setlength{\fboxsep}{1pt}%
\setlength{\fboxrule}{0.4pt}%
\fbox{
\includegraphics[height=2.2cm]{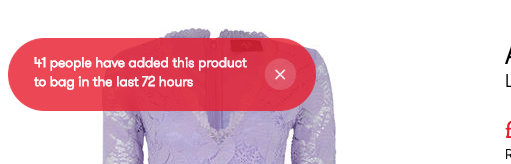}
}
}
}
\hspace*{\fill}
\subfloat[Activity Notification on \texttt{thredup.com}. The message always signals products as if they were sold recently (`just saved'), even in the case of old purchases. \label{fig:sp:an2}]{%
{%
\setlength{\fboxsep}{1pt}%
\setlength{\fboxrule}{0.4pt}%
\fbox{\includegraphics[height=2.2cm]{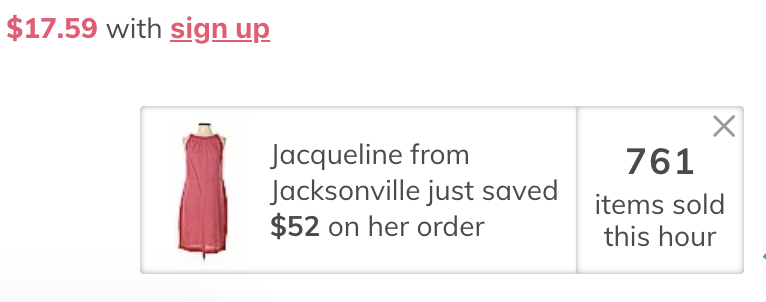}}
}
}

\subfloat[Activity Notification on \texttt{jcpenney.com}. The message indicates the number of people who viewed the product in the 24 hours along with the quantity left in stock.\label{fig:sp:an3}]{
{
\setlength{\fboxsep}{1pt}%
\setlength{\fboxrule}{0.4pt}%
\fbox{
\includegraphics[height=3cm]{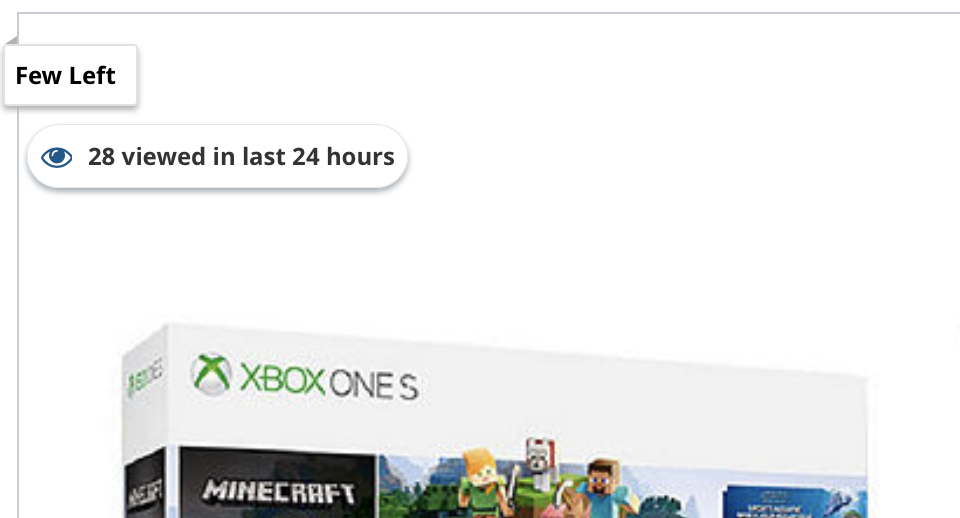}
}
}
}
\quad
\subfloat[Testimonials of Uncertain Origin on \texttt{coolhockey.com}. We found the same testimonials on \texttt{ealerjerseys.com} with different customer names.\label{fig:sp:test}]{%
{%
\setlength{\fboxsep}{1pt}%
\setlength{\fboxrule}{0.4pt}%
\fbox{\includegraphics[height=3cm]{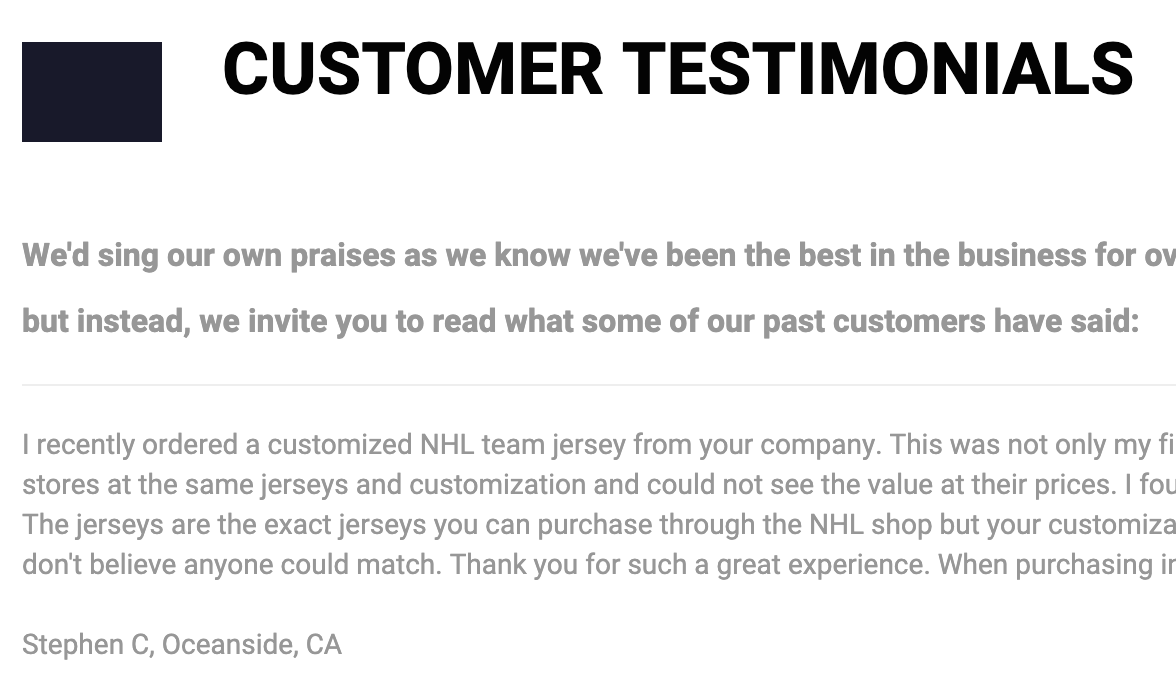}}
}
}

\caption{Two types of the Social Proof category of dark patterns.}
	\label{fig:socialproof}
\end{figure}

We observed two types of the Social Proof dark pattern: Activity Notifications and Testimonials of Uncertain Origin on \spw{} websites across their product and cart pages. In all these instances, the Social Proof messages indicated other users' activities and experiences shopping for products and items. Figure~\ref{fig:socialproof} highlights instances of these two types.

\paragraph{\textbf{Activity Notifications}} The `Activity Notification' dark pattern is a transient, often recurring and attention grabbing message that appears on product pages indicating the activity of other users. These can be grouped into different categories: dynamic and periodic messages that indicated other users just bought a product (e.g., `Abigail from Michigan just bought a new stereo system'); static or dynamic text to indicate how many users have a specific item in their cart (e.g., `35 people added this item to cart'); and similar text to indicate how many users have viewed a product (e.g., `90 people have viewed this product'). Figures~\ref{fig:sp:an1}, \ref{fig:sp:an2}, and~\ref{fig:sp:an3} highlight three instances of Activity Notification on \texttt{tkmaxx.com}, \texttt{thredup.com}, and \texttt{jcpenney.com}, respectively. In our data set, we found a total of \ani{} such instances.

\paragraph{Deceptive Activity Notifications.} We examined the Activity Notification messages in our data set for deceptive practices. To facilitate our analysis, we manually inspected the page source of each shopping website that displayed these notifications to verify their integrity. We ignored all those notifications that were generated server-side since we had limited insight into how and whether they were truly deceptive. We considered an instance of Activity Notification to be deceptive if the content it displayed---including any names, locations statistics, counts---was falsely generated or made misleading statements.

In our data set, we discovered a total of \anid{} instances of deceptive Activity Notifications on \anwd{} shopping websites. The majority of these websites generated their deceptive notifications in a random fashion (e.g., using a random number generator to indicate the number of users who are `currently viewing' a product) and others hard-coded previously generated notifications, meaning they never changed. One notable case was \texttt{thredup.com} as shown in Figure~\ref{fig:sp:an2}, where the website generated messages based on fictitious names and locations for an unvarying list of products that was always indicated to be `just sold'.

Using our taxonomy of dark pattern characteristics, we classify Activity Notifications as partially \emph{covert} (in instances where the notifications are site-wide for example, users may fail to understand their effect on their choices) and sometimes \emph{deceptive} (the content of notifications can be deceptively generated or misleading) in nature.

\paragraph{\textbf{Testimonials of Uncertain Origin}} The `Testimonials of Uncertain Origin' dark pattern refers to the use of customer testimonials whose origin or how they were sourced and created is not clearly specified. For each instance of this dark pattern, we made two attempts to validate its origin. First, we inspected the website to check if it contained a form to submit testimonials. Second, we performed exact searches of the testimonials on a search engine (\texttt{google.com}) to check if they appeared on other websites. Figure~\ref{fig:sp:test} shows one instance on \texttt{coolhockey.com}, where we found the same set of testimonials on \texttt{ealerjerseys.com} with different customer names attached to them. In our data set, we found a total of \testi{} instances of this pattern.

\subsubsection{Scarcity}
\label{section:scarcity}

\begin{figure}[t]
\subfloat[Low-stock Message on \texttt{6pm.com}. Left: Choosing product options shows \emph{Only 3 left in stock}. Right: The out-of-stock product makes it seem that it just sold out.\label{fig:scarcity:ls1}]{
{
\setlength{\fboxsep}{3pt}%
\setlength{\fboxrule}{0.4pt}%
\fbox{
\includegraphics[height=2cm]{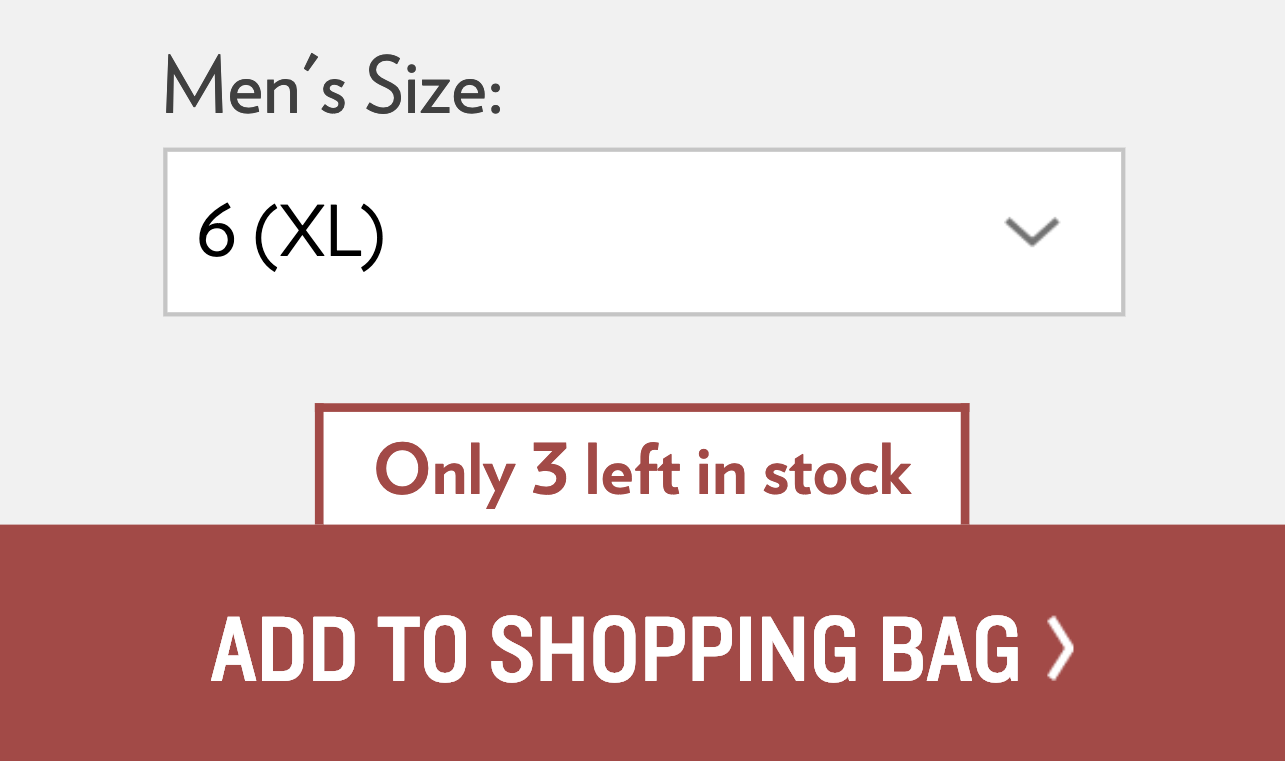}
\includegraphics[height=1.5cm]{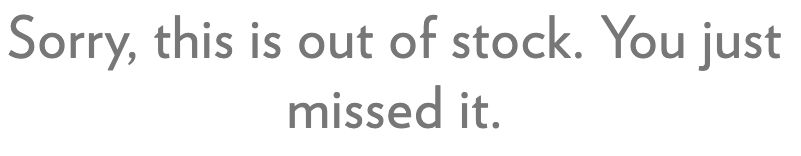}
}
}
}

\subfloat[Low-stock on \texttt{orthofeet .com}. Appears for all products.  \label{fig:scarcity:ls2}]{%
{%
\setlength{\fboxsep}{3.2pt}%
\setlength{\fboxrule}{0.4pt}%
\fbox{\includegraphics[height=2cm]{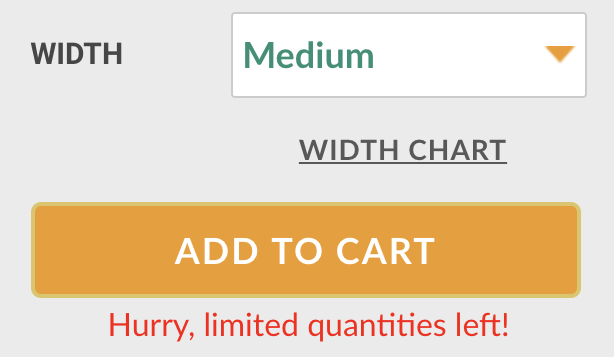}}
}
}
\hspace{12pt}
\subfloat[High-demand Message on \texttt{fashionnova.com}. The message appears for all products in the cart. \label{fig:scarcity:hd}]{%
{%
\setlength{\fboxsep}{3.2pt}%
\setlength{\fboxrule}{0.4pt}%
\fbox{\includegraphics[height=2cm]{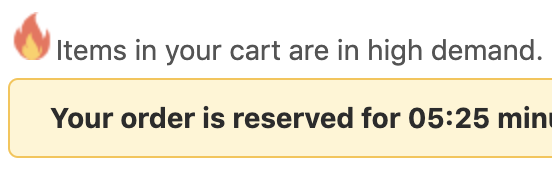}}
}
}

\caption{Two types of the Scarcity category of dark patterns.}
	\label{fig:scarcity}
\end{figure}

`Scarcity' refers to the category of dark patterns that signal the limited availability or high demand of a product, thus increasing its perceived value and desirability~\cite{lynn1991scarcity,scarcity-jung, cialdini2009influence,nodder2013evil}. We observed two types of the Scarcity dark pattern: `Low-stock Messages' and `High-demand Messages' on \scarw{} shopping websites across their product and cart pages. In both pages, they indicated the limited availability of a product or that a product was in high demand and thus likely to become unavailable soon. Figure~\ref{fig:scarcity} highlights instances of these two types.

\paragraph{\textbf{Low-stock Messages}} The `Low-stock Message' dark pattern signals to users about limited quantities of a product. Figure~\ref{fig:scarcity:ls1} shows an instance of this pattern on \texttt{6pm.com}, displaying the precise quantity in stock. In our data set, we found a total of \lsi{} instances of the Low-stock Message dark pattern. However, not all of these instances displayed stock quantities. 49 of these instances only indicated that stock was limited or low, without displaying the exact quantity, resulting in uncertainty, increased desirability of products, and impulse buying behavior in users. Figure~\ref{fig:scarcity:ls2} shows one such instance on \texttt{orthofeet.com}.

\paragraph{Deceptive Low-stock Messages.} We examined all the Low-stock Message dark patterns for deceptive practices using the method described in Section~\ref{section:detecting-deceptive-adp}. From the resulting data, we ignored those websites whose stock amounts remained the same between visits, reasoning that those are unlikely to be indicative of deceptive practices. We then manually examined the remaining sites and identified how the stock information was generated.

In our data set, we discovered a total of \lsdi{} instances of deceptive Low-stock Messages on \lsdw{} shopping websites. On further examination, we observed that 16 of these sites decremented stock amounts in a recurring, deterministic pattern according to a schedule, and the one remaining site (\texttt{forwardrevive.com}) randomly generated stock values on page load. Exactly 8 of these sites used third-party JavaScript libraries to generate the stock values, such as Hurrify~\cite{hurrify} and Booster~\cite{boostify}. Both of these are popular plugins for Shopify---one of the largest e-Commerce companies---based websites. The remaining websites injected stock amounts through first-party JavaScript or HTML.

Besides the use---or non-use---of numeric data and deception, Low-stock Messages can be concerning in other ways. For example, we observed that several websites, such as \texttt{6pm.com} and \texttt{orthofeet.com}, displayed Low-stock Messages for nearly all their products---stating `Only X left' and `Hurry, limited quantities left!' respectively. The former, in particular, showed a `Sorry, this is out of stock. You just missed it' popup dialog for every product that was sold out, even if it had already been out of stock in the previous days.

Using our taxonomy of dark pattern characteristics, we classify Low-stock Messages as partially \emph{covert} (it creates a heightened sense of impulse buying, unbeknownst to users), sometimes \emph{deceptive} (it can mislead users into believing a product is low on stock when in reality it is not, creating false scarcity), and sometimes \emph{information hiding} (in some instances, it does not explicitly specify the stock quantities at hand) in nature.

\paragraph{\textbf{High-demand Messages}} The `High-demand Message' dark pattern signals to users that a product is in high demand, implying that it is likely to sell out soon. Figure~\ref{fig:scarcity:hd} shows one such instance on \texttt{fashionnova.com} on the cart page, indicating that the products in the cart are selling out quickly. In our data set, we found a total of \hdi{} instances of the High-demand dark pattern; 38 of these instances appeared consistently, regardless of the product displayed on the website, or regardless of the items in cart. As with Low-stock Messages, we classify High-demand Messages as partially \emph{covert}.



\subsubsection{Obstruction}
\label{section:obstruction}

`Obstruction', coined by Gray et al.\ ~\cite{gray-dark-patterns-2018}, refers to the category of dark patterns that make a certain action harder than it should be in order to dissuade users from taking that action. We observed one type of the Obstruction dark pattern: `Hard to Cancel'---a pattern similar to Brignull's \emph{Roach Motel} dark pattern~\cite{brignull-dark-patterns}---on \htcw{} websites. Obstruction makes it easy for users to sign up for recurring subscriptions and memberships, but it makes it hard for them to subsequently cancel the subscriptions.

\begin{figure}[t]

\centering
\subfloat[Hard to Cancel on \texttt{sportsmanguide.com}. The website only discloses in the terms PDF file that canceling the recurring service requires calling customer service.\label{fig:o2}]{%
{%
\setlength{\fboxsep}{1pt}%
\setlength{\fboxrule}{0.4pt}%
\fbox{\includegraphics[height=0.8cm]{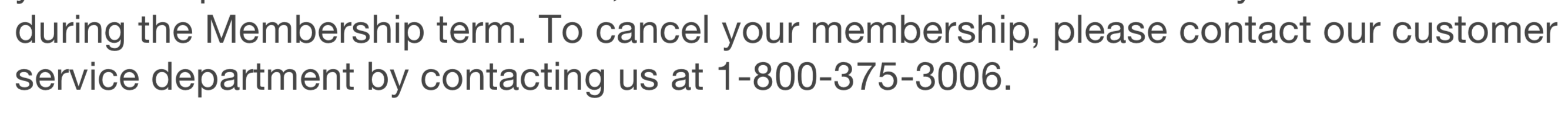}
}
}
}

\subfloat[Hard to Cancel on \texttt{savagex.com}. The website discloses upfront that the recurring service can only be canceled through customer care. \label{fig:o1}]{
{
\setlength{\fboxsep}{1pt}%
\setlength{\fboxrule}{0.4pt}%
\fbox{
\includegraphics[height=2.cm]{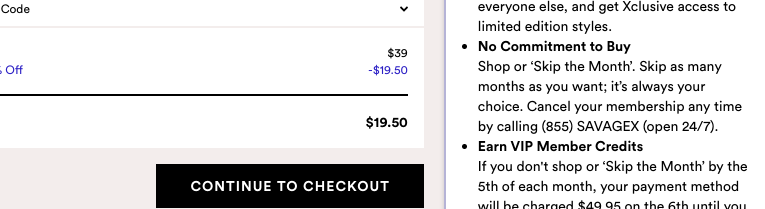}
}
}
}

\caption{The Hard to Cancel type from the Obstruction category of dark patterns.}
	\label{fig:obstruction}
\end{figure}

More often than not, shopping websites did not disclose upfront to users that canceling the subscription or membership could not be completed in the same manner they signed up for the memberships in the first place. For example, as shown in Figure~\ref{fig:o2}, \texttt{sportsmansguide.com} promotes a `buyer's club' discount membership price and makes it easy for users to sign up for the annual recurring membership, as they are under the impression they can `cancel anytime.' However, \texttt{sportsmansguide.com}'s terms of service reveal that the membership can only be cancelled by calling their customer service. In rare instances, as shown in Figure~\ref{fig:o1}, websites such as \texttt{savagex.com} disclosed upfront that cancellation required calling customer service.

Using our taxonomy of dark pattern characteristics, we classify Hard to Cancel as \emph{restrictive} (it limits the choices users can exercise to cancel their services) in nature. In cases where websites do not disclose their cancellation policies upfront, Hard to Cancel also becomes \emph{information hiding} (it fails to inform users about how cancellation is harder than signing up) in nature.

\subsubsection{Forced Action}

`Forced Action' refers to the category of dark patterns---originally proposed by Gray et al.\ ~\cite{gray-dark-patterns-2018}---that require users to take certain additional and tangential actions to complete their tasks. We observed one type of the Forced Action dark pattern, `Forced Enrollment', on \few{} websites. This type of dark pattern explicitly coerces users into signing up for marketing communication, or creates accounts to surrender users' information. By using the Forced Enrollment dark pattern, online services and websites collected more information about their users than they might otherwise consent to---resulting from an all-or-nothing proposition.

\begin{figure}[t]
\centering
\subfloat[Forced Enrollment on \texttt{musiciansfriend.com}. Agreeing to the terms of use also requires agreeing to receive emails and promotions.\label{fig:fa2}]{%
{%
\setlength{\fboxsep}{1pt}%
\setlength{\fboxrule}{0.4pt}%
\fbox{\includegraphics[height=3.7cm]{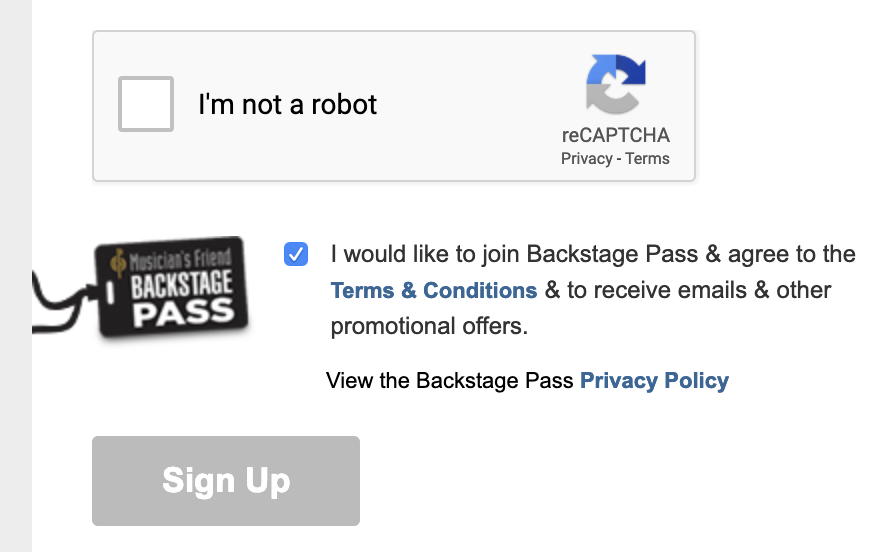}
}
}
}
\hspace*{\fill}
\subfloat[Forced Enrollment on \texttt{therealreal.com}. Browsing the website requires creating an account even without making a purchase. \label{fig:fa1}]{
{
\setlength{\fboxsep}{1pt}%
\setlength{\fboxrule}{0.4pt}%
\fbox{
\includegraphics[height=3.7cm]{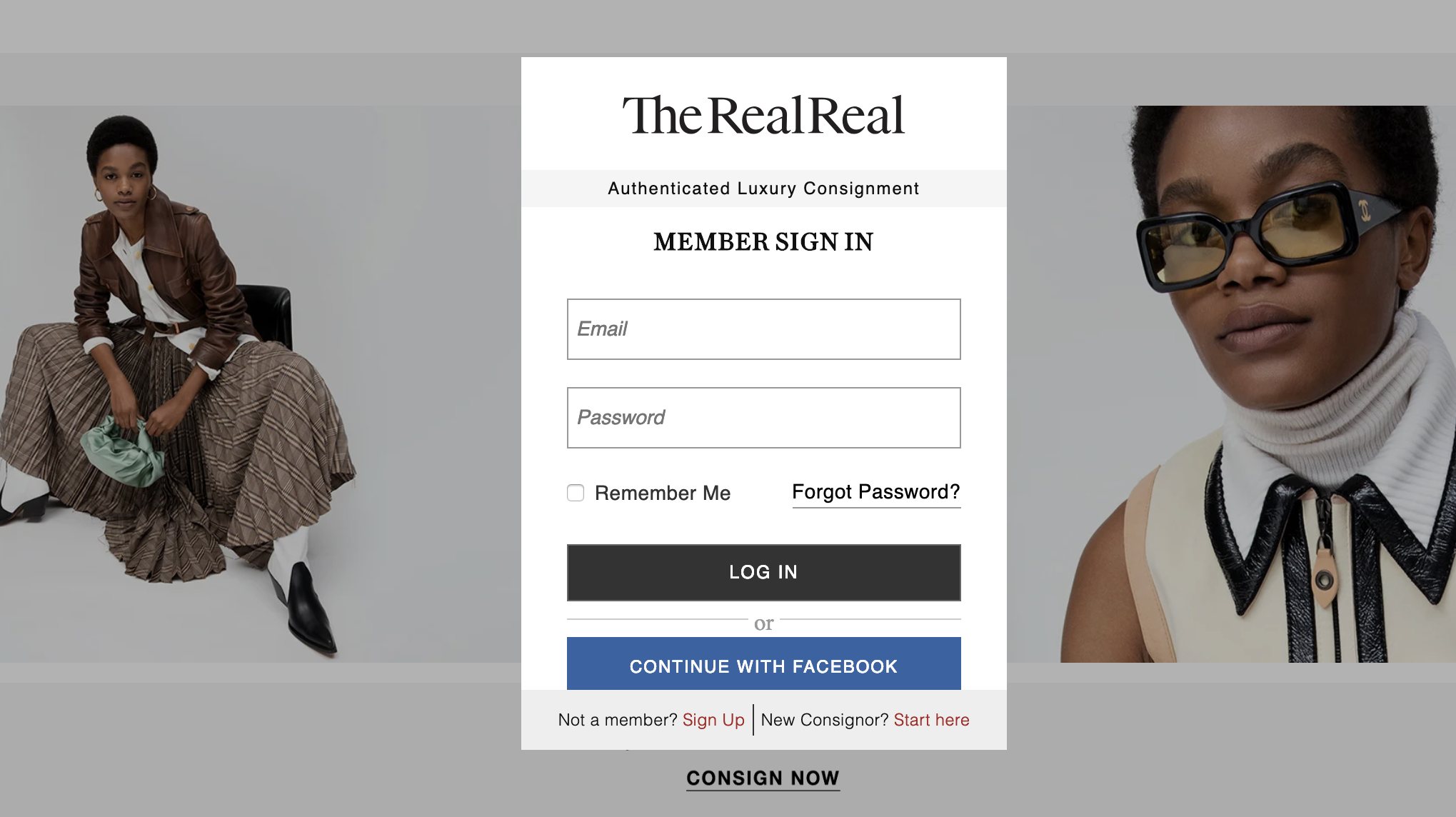}
}
}
}

\caption{The Forced Enrollment type from the Forced Action category of dark patterns.}
	\label{fig:forced-action}
\end{figure}

On four out of six websites, the Forced Enrollment dark pattern manifested as a checkbox in the user interface, requiring users to simultaneously consent to the terms of service \emph{and} to receiving marketing emails as part of the consent process. Figure~\ref{fig:fa2} shows one such instance on \texttt{musiciansfriend.com}. In another instance of the Forced Enrollment on \texttt{therealreal.com}---as shown in Figure~\ref{fig:fa1}---the website displayed a popup dialog that prevented users from viewing product offerings on the website without creating an account---even if users eventually decide against making a purchase.

Using our taxonomy of dark pattern characteristics, we classify Forced Enrollment as \emph{asymmetric} (it requires competing the additional, tangential tasks, creating unequal choices) and \emph{restrictive} (it mandates enrolling in marketing communication or creating accounts) in nature.

\subsection{Dark Patterns as A Third-Party Service: A Case Study Of Social Proof Activity Notifications} \label{section:entities-that-enable-adversarial-design-patterns}

In many instances, third-party entities---i.e., organizations and companies other than the shopping websites themselves---were responsible for creating and presenting dark patterns on behalf of the shopping websites. We observed this frequently to be the case for one dark pattern in particular: Social Proof Activity Notifications (Section~\ref{section:findings:social-proof}). In this section, we shed light on the ecosystem of third parties that enable Social Proof Activity Notifications, using our starting point as the list of websites in our data set that displayed such Activity Notifications.

\subsubsection{Detecting Third-party Entities}

In order to detect third-party entities, it is sufficient to uncover scripts that are served from third-party domains and are responsible for creating Social Proof Activity Notifications. However, automatically attributing certain interface elements and webpage modifications to third-party scripts constitutes a more challenging task because modern browsers do not expose any means to attribute DOM changes (e.g. displaying a popup dialog) to particular scripts. Further, web pages may be modified by several different first and third-party scripts in the same visit, making attribution trickier.


To overcome this challenge, we employed a combination of automated and manual analyses. We used the following observation: when a third-party entity displays an Activity Notification on a shopping website, its content should be included in the HTTP response received from this third party's servers on that website. For example, if the notification states `Jane from Washington, DC just purchased this product', looking up the customer name and location---in this case `Jane' and `Washington, DC'---in the HAR file for that website should reveal the end point of the server that issued the notification. Thus, for all notifications of this kind, we extracted the name and location pairs from the content, searched the HAR files for these pairs; where successful, we recorded the HTTP endpoints corresponding to the third-parties. We then manually verified these endpoints and determined the responsible entities by using the WHOIS database, visiting the script domains and using search engines to uncover the company identities and websites.

Where this analysis failed to return an HTTP endpoint from the HAR files, and for all other kinds of Social Proof Activity Notification (e.g., `This product was added to cart 10 times in the last day'), we manually visited the websites containing the message to determine the responsible third parties. We sped up this analysis using Google Chrome Developer Tool's `DOM change breakpoints' feature~\cite{chrome-dev-tools-pause}, which helped us easily determine the responsible entities.

Having determined the third-party entities, we measured their prevalence across all the shopping websites in our data set. To do so, we searched the HTTP request data from checkout crawls for the third-party domains we identified. Finally, as a reference point, we also determined their prevalence on the web---beyond shopping websites---using the latest publicly available crawl data (November 2018) from the Princeton Web Census Project~\cite{princeton-webcensus, englehardt-tracking-2017}. This public project documents the prevalence of third-party scripts using periodic scans of home pages of Alexa top million sites and is available for external researchers to use. 


\subsubsection{The Ecosystem Of Third-party Entities}

\begin{table}[]
\scriptsize
\caption{List and prevalence of Social Proof Activity Notifications enabling third-party entities in our data set of 11K shopping websites and the home pages of Alexa top million websites~\cite{princeton-webcensus}. Where available, we list additional dark patterns the third parties claim to offer. Nice/Bizzy, Woocommerce Notification, Boost, and Amasty are Shopify, Woocommerce, Wordpress and Magento plugins respectively.}
\begin{tabular}{lrrl}
\multirow{2}{*}{\textbf{\begin{tabular}[c]{@{}c@{}}Third-party\\ Entity\end{tabular}}} & \multicolumn{2}{c}{\textbf{Prevalence}} & \multirow{2}{*}{\textbf{\begin{tabular}[c]{@{}c@{}}Additional Dark\\ Patterns\end{tabular}}} \\ \cline{2-3}
 & \textbf{\begin{tabular}[c]{@{}c@{}}\# Shopping\\ Websites\end{tabular}} & \multicolumn{1}{c}{\textbf{\begin{tabular}[c]{@{}c@{}}\# Alexa Top \\ Million\end{tabular}}} &  \\ \hline

Beeketing & 406 & 4,151 & Pressured Selling, Urgency, Scarcity \\
Dynamic Yield & 114 & 416 & Urgency \\
Yieldify & 111 & 323 & Urgency, Scarcity \\ 
Fomo & 91 & 663 & -- \\
Fresh Relevance & 86 & 208 & Urgency\\
Insider & 52 & 484 & Scarcity, Urgency \\
Bizzy &  33 & 213 & -- \\
ConvertCart & 31 & 62 & -- \\
Taggstar & 27 & 4 & Scarcity, Urgency \\
Qubit & 25 & 73 & Pressured Selling, Scarcity, Urgency \\
Exponea & 18 & 180 & Urgency, Scarcity \\
Recently & 14 & 66 & -- \\
Proof & 11 & 508 & -- \\
Fera & 11 & 132 & Pressured Selling, Scarcity, Urgency \\
Nice & 10 & 80 & -- \\
Woocommerce Notification & 10 & 61 & -- \\
Bunting & 5 & 17 & Urgency, Scarcity \\
Credibly & 4 & 67 & -- \\
Convertize & 3 & 58 & Scarcity, Urgency \\
LeanConvert & 2 & 0 & -- \\
Boost & 1 & 3 & -- \\
Amasty  & 1 & 0 & Pressured Selling, Scarcity, Urgency \\ \hline
\end{tabular}
\label{tab:tps}
\end{table}

Table~\ref{tab:tps} summarizes our findings. We discovered a total of 22 third-party entities, embedded in 1,066 of the 11K shopping websites in our data set, and in 7,769 of the Alexa top million websites. We note that the prevalence figures from the Princeton Web Census Project data should be taken as a lower bound since their crawls are limited to home pages of websites. This difference in prevalence is particularly visible for certain third-party entities like Qubit and Taggstar, where their prevalence is higher in our data set compared to the Web Census data. By manually examining websites that contained these third parties, we discovered that many shopping websites only embedded them in their product---and not home---pages, presumably for functionality and performance reasons.

We learned that many third-party entities offered a variety of services for shopping websites, including plugins for popular e-commerce platforms such as Shopify\footnote{https://shopify.com} and Woocommerce\footnote{https://woocommerce.com}. To better understand the nature and capabilities of each third-party entity, we examined any publicly available marketing materials on their websites. 

Broadly, we could classify the third-party entities into two groups. The first group exclusively provided Social Proof Activity Notifications integration as a service. The second group provided a wider array of marketing services that often enabled other types of dark patterns; most commonly these were Scarcity and Urgency dark patterns. We list all these additional dark pattern capabilities in the rightmost column of Table \ref{tab:tps}.


Many of the third-parties advertised practices that appeared to be---and sometimes unambiguously were---manipulative: `[p]lay upon [customers'] fear of missing out by showing shoppers which products are creating a buzz on your website' (Fresh Relevance), `[c]reate a sense of urgency to boost conversions and speed up sales cycles with Price Alert Web Push' (Insider), `[t]ake advantage of impulse purchases or encourage visitors over shipping thresholds' (Qubit). Further, Qubit also advertised Social Proof Activity Notifications that could be tailored to users' preferences and backgrounds. 

In some instances, we found that third parties openly advertised the deceptive capabilities of their products. For example, Boost dedicated a web page---titled `Fake it till you make it'---to describing how it could help create fake orders~\cite{boost-fake}. Woocommerce Notification---a Woocommerce platform plugin---also advertised that it could create fake social proof messages: `[t]he plugin will create fake orders of the selected products'~\cite{woo-commerce-notification}. Interestingly, certain third parties (Fomo, Proof, and Boost) used Activity Notifications on their websites to promote their own products.

Finally, we also discovered that some of these deceptive practices resulted in e-commerce platforms taking action against third-party entities. For instance, Beeketing's---the most popular third party provider in our data set---`Sales Pop' Shopify plugin was temporarily removed from Shopify in an effort to crack down on deceptive practices~\cite{beeketing-removed, beeketing-removed-FB-official}. The plugin had allowed websites to create fake Activity Notifications by entering fabricated sales data.

In summary, we discovered that third party entities widely enable dark patterns on shopping websites. Furthermore, some of these third-parties even advertised the deceptive use of their services.

\section{Discussion}
\label{section:discussion}
\begin{figure}[t]
\centering

\frame{\includegraphics[width=0.7\textwidth]{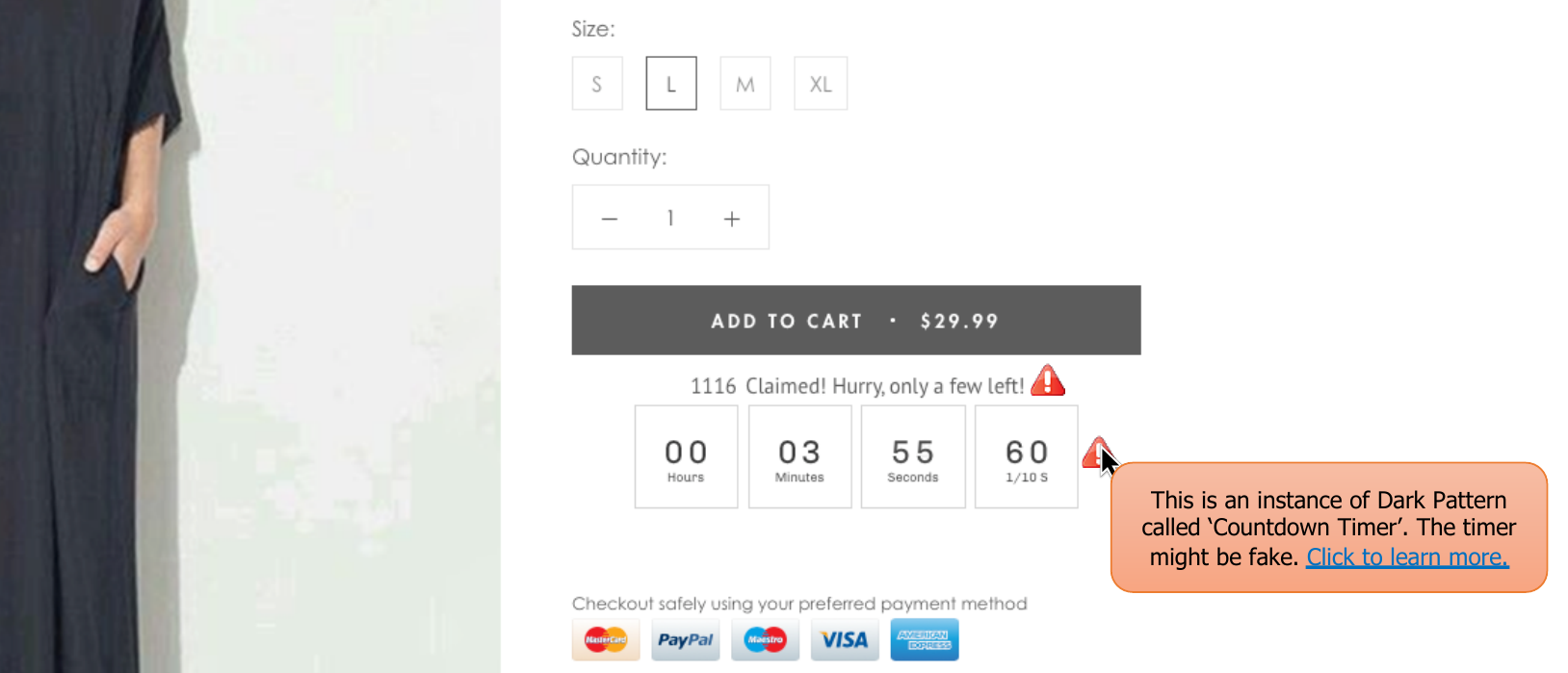}}
\caption{Mockup of a possible browser extension that can be developed using our data set. The extension flags instances of dark patterns with a red warning icon. By hovering over the icon, the user can learn more about the specific pattern.}
\label{fig:browser-extension-mockup}
\end{figure}

\subsection{Dark Patterns and Implications For Consumers}
Many dark patterns constitute manipulative and deceptive practices that past work has shown users are increasingly becoming aware of~\cite{Chivukula:2019:CBP:3290607.3312863}. Our current data set of dark patterns, comprising of screenshots and text segments, can be used to build countermeasures to help users make more informed decisions even in the presence of dark patterns. One such countermeasure could be a public-facing website that scores shopping websites based on their use of dark patterns. Our data set can also enable the development of browser extensions that automatically detect and flag dark patterns (e.g., shopping websites, as shown in Figure \ref{fig:browser-extension-mockup}). Such a tool could be augmented to flag dark patterns on websites not in our data set through users' submissions, through community-generated and maintained lists (similar to how ad blockers work~\cite{adBlock}), or through trained machine learning classifiers. Eventually, such tools could be integrated into browsers themselves. For example, in recent years, Firefox and Safari have shown interest in integrating tools that promote consumer privacy (e.g., features to block web tracking by default~\cite{webkitIntelligentTrackingProtection,mozillaTrackingProtection}). However, finding the right incentives for browser vendors to implement these solutions might be challenging in the context of dark patterns, since they might be wary of policing content on the web. Finally, future studies could leverage our descriptive and comparative taxonomy of dark pattern characteristics to better understand their effects on users, as well as to ascertain which dark patterns are considered most egregious by users (e.g., by means of users studies).

\subsection{Implications for Consumer Protection Policy and Retailers}
Our results demonstrate that a number of shopping websites use deceptive dark patterns, involving affirmative and false representations to consumers. 
We also found 22 different third-party entities that enable the creation of Social Proof Activity Notification dark patterns. Some of these entities promote blatantly deceptive practices and provide the infrastructure for retailers to use these practices to influence consumer behavior for profit. These practices are unambiguously unlawful in the United States (under Section 5 of the Federal Trade Commission Act and similar state laws \cite{ftcact-sec5}), and the European Union (under the Unfair Commercial Practices Directive and similar member state laws \cite{eu-2011-83}).

We also find practices that are unlawful in a smaller set of jurisdictions. In the European Union, businesses are bound by an array of affirmative disclosure and independent consent requirements in the Consumer Rights Directive \cite{eu-inertia-selling}. Websites that use the Sneaking dark patterns (Sneak into Basket, Hidden Subscription, and Hidden Costs) on European Union consumers are likely in violation of the Directive. Furthermore, user consent obtained through Trick Questions and Visual Interference dark patterns do not constitute freely given, informed and active consent as required by the General Data Protection Regulation (GDPR)~\cite{gdpr}. In fact, the Norwegian Consumer Council filed a GDPR complaint against Google in 2018, arguing that Google used dark patterns to manipulate users into turning on the `Location History' feature on Android, and thus enabling constant location tracking~\cite{gdpr-complaint}.

In addition to demonstrating specific instances of unlawful business practices, we contribute a new approach for regulatory agencies and other consumer protection stakeholders (e.g., journalists and civil society groups) to detect dark patterns. The crawling and clustering methodology that we developed is readily generalizable, and it radically reduces the difficulty of discovering and measuring dark patterns at web scale. Furthermore, our data set of third-party entities which provide the infrastructure to enable certain deceptive dark patterns can be used by regulators as a starting point to inform policy and regulation around what kinds of practices should be allowable in the context online shopping.

\subsection{Dark Patterns and Future Studies At Scale}
We created automated techniques that can be used to conduct measurements of dark patterns at web scale. Researchers can extend our tools and infrastructure to document the presence of dark patterns other types of websites (e.g., travel and ticket booking websites) by building a crawler that traverses users' primary interaction paths on those websites. Researchers can also extend our techniques to measure dark patterns that are not inherently \emph{dark} because of the text they display but because they take advantage of visual elements. For example, urgency can be created by a blinking timer; similarly, Hidden Subscriptions can make the default option (e.g., subscribing to a paid service) visually more appealing and noticeable than its alternative (e.g., not subscribing). One starting point to detect such interfaces could be to incorporate style and color as features for clustering, or even use the design mining literature~\cite{kumar-webzeitgeist,kumar-design-semantics, kumar-rico} to analyze specific types of interfaces (e.g., page headers) in isolation. Finally, researchers can leverage our descriptive taxonomy of dark pattern characteristics to study and analyze dark patterns in other domains, such as emails and mobile applications.

\subsection{Limitations}
Our research has several limitations. First, we only take into account text-based dark patterns and, therefore, leave out those that are inherently visual (e.g., using font size or color to emphasize one part of the text more than another). Second, many of the dark patterns we document are derived from the existing dark patterns literature. However, some of these are exist in a gray area, and in those cases determining whether a dark pattern is deliberately misleading or not can sometimes be hard to discern. Opinions of dark patterns may also vary between and among experts and users (e.g., countdown timers to indicate when to order to be eligible for free shipping). Clarifying this gray area and establishing the degree to which these patterns are perceived as manipulative by users can be further investigated by future user studies. Third, in Section \ref{section:definitions} we drew connections between each type of dark pattern and a set of cognitive biases it exploits. However, these connections may be more nuanced or complex. For example, not all individuals may be equally susceptible to these cognitive biases; some individuals may be more susceptible to one kind over another. Fourth, during our crawls we experienced a small number of Selenium crashes, which did not allow us to either retrieve product pages or complete data collection on certain websites. Fifth, while the crawler was mostly effective in simulating user actions, it failed to complete the product purchase flow on some websites (see Section~\ref{section:method}). Sixth, and finally, we only crawled product pages and checkout pages, missing out on dark patterns commonly present in other pages, such as the home page, product search, and account creation pages. Many dark patterns also appear after purchase (e.g., upselling) which our crawler fails to capture because we do not make purchases. Future studies could consider collecting these kinds of dark patterns from users.


\section{Conclusion}
\label{section:conclusion}
In this paper, we developed automated techniques to study dark patterns on the web at scale. By simulating user actions on the $\sim$11K most popular shopping websites, we collected text and screenshots of these websites to identify their use of dark patterns. We defined and characterized these dark patterns, describing how they affect users' decisions by linking our definitions to the cognitive biases leveraged by dark patterns. We found at least one instance of dark pattern on approximately \totaldpperc{} of the examined websites. Notably, \totaldecw{} of the websites displayed deceptive messages. Furthermore, we observed that dark patterns are more likely to appear on popular websites. Finally, we discovered that dark patterns are often enabled by third-party entities, of which we identify 22; two of these advertise practices that enable deceptive patterns. Based on these findings, we suggest that future work focuses on empirically evaluating the effects of dark patterns on user behavior, developing countermeasures against dark patterns so that users have a fair and transparent experience, and extending our work to discover dark patterns in other domains.

\begin{acks}
We are grateful to Mihir Kshirsagar, Finn Myrstad, Vincent Toubiana, and Joe Calandrino for feedback on this paper.
\end{acks}

\bibliographystyle{ACM-Reference-Format}
\bibliography{main}


\pagebreak
\appendix
\section{Appendix}
\vspace{5.5mm}

\begin{table}[h]
\caption{Confusion Matrices From Our Evaluation of Alexa's and Webshrinker's Website Classifiers.}
\begin{tabular}{cc|cc|cc}
\multicolumn{1}{c}{} &\multicolumn{1}{c}{} &\multicolumn{2}{c|}{Alexa Prediction} &\multicolumn{2}{c}{Webshrinker Prediction}\\ 
\cline{3-6}
\multicolumn{1}{c}{} & 
\multicolumn{1}{c}{} & 
\thead{Not Shopping} & 
\thead{Shopping} &
\thead{Not Shopping} & 
\thead{Shopping} \\ 
\cline{2-6}
\multirow[c]{2}{*}{\rotatebox[origin=tr]{90}{Truth}}
& \thead{Not Shopping}  & 442 & 1 & 423 & 20  \\[1.5ex]
& \thead{Shopping}  & 53   & 4 & 10 & 47\\
\cline{2-6}
\end{tabular}
\label{tab:confusion-matrices}
\end{table}

\begin{figure}[h]
\centering
\vspace{15mm}
\resizebox{\textwidth}{!}{%
\frame{\includegraphics[width=\linewidth]{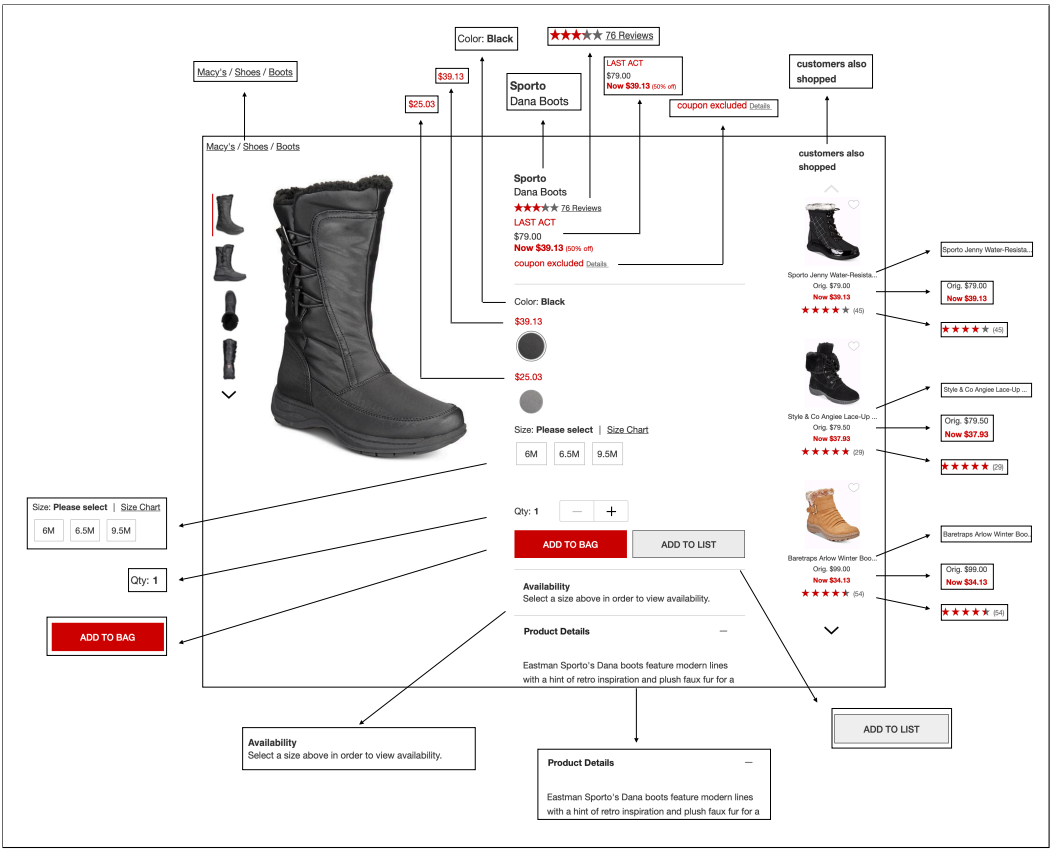}}}
\caption{An illustration of the page segmentation algorithm. The page is segmented into smaller meaningful ``building blocks'' or segments. Only segments containing text are recorded.}
\label{fig:segmentation}
\end{figure}

\begin{algorithm}
  \caption{Page Segmentation}
  \label{alg:segments}
  \begin{algorithmic}[1]
\State $ignoredElements \gets$ [`script', `style', `noscript', `br', `hr']
\State $blockElements \gets$ [`div', `section', `article', `aside', `nav',
 `header', `footer', `main', `form', `fieldset', `table']
\State
\Function{segments}{element}
\Comment{Returns a list of segments}
\If {not $element$}
    \ReturnNewLine empty list
\EndIf
\State $tag\gets$ $element.tagName$
\If {$tag$ in $ignoredElements$ or $element$ not visible or $element$ not bigger than 1 pixel}
	\ReturnNewLine empty list
\EndIf
\If{$tag$ in $blockElements$}
	\If {$element$ does not contain visible $blockElements$}
		\If {all of $element$'s children in $ignoredElements$}
			\ReturnNewLine empty list
		\Else
			\If {$element$ occupies more than 30\% of the page}
				\ReturnNewLine list of $segments(child)$ for each child in $element$'s children
			\Else
			    \ReturnNewLine [$element$]
			\EndIf
		\EndIf
	\ElsIf{$element$ contains text nodes}
		\ReturnNewLine [$element$]
	\Else
		\ReturnNewLine list of $segments(child)$ for each child in $element$'s children
	\EndIf
\Else
	\If {$element$ has at least one child with $tag$ in $blockElements$}
		\ReturnNewLine list of $segments(child)$ for each child in $element$'s children
	\Else
		\If{$element$ occupies more than 30\% of the page}
			\ReturnNewLine list of $segments(child)$ for each child in $element$'s children
		\Else \ReturnNewLine [$element$]
		\EndIf
	\EndIf
\EndIf

\EndFunction
\end{algorithmic}
\end{algorithm}

\received{April 2019} 
\received[revised]{June 2019}
\received[accepted]{August 2019}

\end{document}